\documentclass[11pt]{amsart}
\usepackage{amsbsy,amssymb,amscd,amsfonts,latexsym,amstext,delarray,
amsmath,graphicx} 

\newtheorem{thm}{Theorem}[section]
\newtheorem{prop}[thm]{Proposition}
\newtheorem{cor}[thm]{Corollary}
\newtheorem{lem}[thm]{Lemma}

\newtheorem{rem}[thm]{Remark}

\numberwithin{equation}{section}

\def\C{{\mathbb C}}

\def\N{{\mathbb N}}

\def\Z{{\mathbb Z}}
\def\R{{\mathbb R}}

\def\cA{{\mathcal A}}

\def\cD{{\mathcal D}}

\def\cH{{\mathcal H}}

\def\cL{{\mathcal L}}

\def\cS{{\mathcal S}}

\def\cV{{\mathcal V}}
\def\cW{{\mathcal W}}

\def\cZ{{\mathcal Z}}

\def\Ker{{\rm Ker}}

\def\Spec{{\rm Spec}}
\def\Sp{{\rm Spec}}

\def\Tr{{\rm Tr}}

\title{The spectral action and cosmic topology}
\author{Matilde Marcolli}
\author{Elena Pierpaoli}
\author{Kevin Teh}
\address{Department of Mathematics  \\
California Institute of Technology \\ 
Pasadena, CA 91125, USA}
\email{matilde\@@caltech.edu}
\email{teh\@@caltech.edu}
\address{Department of Physics and Astronomy \\
University of Southern California \\
Los Angeles, CA 90089, USA}
\email{pierpaol\@@usc.edu}

\begin{document}
\maketitle

\begin{verse}
{\em In memory of Andrew Lange}
\end{verse}

\begin{abstract}
The spectral action functional, considered as a model of gravity coupled to matter,
provides, in its non-perturbative form, a slow-roll potential for inflation, whose form
and corresponding slow-roll parameters can be sensitive to the underlying 
cosmic topology. We explicitly compute the non-perturbative spectral action for
some of the main candidates for cosmic topologies, namely the quaternionic
space, the Poincar\'e dodecahedral space, and the flat tori. We compute the 
corresponding slow-roll parameters and see we check that the resulting inflation 
model behaves in the same way as for a simply-connected spherical topology in 
the case of the quaternionic space and the Poincar\'e homology sphere, while it 
behaves differently in the case of the flat tori. We add an appendix with
a discussion of the case of lens spaces.
\end{abstract}

\tableofcontents

\section{Introduction}

Noncommutative geometry provides models of particle physics, with matter Lagrangians
coupled to gravity, based on an underlying geometry which is a product of an ordinary
4-dimensional (commutative) spacetime manifold by a small noncommutative space
which determines the matter content of the model. The spectral action functional, which
is defined for metric noncommutative spaces (spectral triples) is obtained as the trace of
a cutoff of the Dirac operator of the spectral triple by a test function. The asymptotic
expansion of the spectral action delivers a classical Lagrangian, which contains
gravitational terms (Einstein--Hilbert, conformal gravity, cosmological term) and
a coupled matter Lagrangian. For a suitable choice of the noncommutative space
the latter recovers the Standard Model Lagrangian and some extensions with right handed
neutrinos, and more recently with supersymmetric QCD, see \cite{CCM}, \cite{BroSu}.

\smallskip

In trying to understand the cosmological implications of this model, one can work as
in \cite{MaPie} with the asymptotic expansion of the spectral action functional, but 
this only delivers models of the very early universe, near the unification epoch. These
can be potentially interesting, as the model contains different possible mechanisms
of inflation, related to the presence of effective gravitational and cosmological
constants. However, one cannot extrapolate that form of the model towards the modern
universe, due to the possible presence of non-pertubative effects in the spectral action
at lower energies. The spectral action in its non-perturbative form is typically very difficult
to compute exactly. However, the recent result of \cite{uncanny} shows that, for a
spacetime whose spatial sections are 3-spheres $S^3$, Wick rotated and
compactified to a Euclidean model $S^3 \times S^1$, the spectral action can be
computed explicitly in non-perturbative form, through a careful use of the Poisson 
summation formula. 

\smallskip

In particular, we show here that the non-perturbative correction observed in
\cite{uncanny} for perturbations $D^2 \mapsto D^2 + \phi^2$ of the Dirac operator
gives rise to a slow-roll potential $V(\phi)$ for the field $\phi$, which can be used
as a model for inflation. We compute the corresponding slow-roll parameters.
These are independent of the artifact of the Euclidean compactification to $S^3\times S^1$.
In particular the dependence on the $\beta$ parameter coming from the size of the
$S^1$ factor disappears. Moreover, while in the Euclidean compactification, the
energy scale $\Lambda$ and the sphere radius $a$ are independent quantities, for
a Lorentzian geometry with the Friedmann form of the metric, both
the scale factor $a(t)$ and the energy scale $\Lambda(t)$ become time-dependent
quantities, related by $\Lambda =1/a(t)$. Since in the explicit nonperturbative
form of the spectral action only the product $\Lambda a$ appears, the resulting
slow-roll potential continues to make sense in the Lorentzian signature, and
the dependence on the scale factor $a(t)$ disappears through being matched with the
$\Lambda =1/a(t)$ scale. Thus, the slow-roll mechanism obtained from the
non-perturbative form of the spectral action can be Wick rotated back to the
Lorentzian Friedmann form of the geometry and used as a model from which
to derive estimates on cosmological parameters such as the spectral index $n_s$
and the tensor-to-scalar ratio.

\smallskip

Thus, we obtain a slow-roll potential for inflation from the non-perturbative
corrections to the spectral action, which is potentially sensitive to the geometry
and topology of the underlying 3-dimensional sections of spacetime. This is
interesting, in the perspective of deriving cosmological signatures of 
possibly non-simply connected topologies. This is known as the problem
of cosmic topology and it has been widely studied by cosmologists in
recent years. We review briefly the current state of understanding of this
problem and the list of those that are currently considered to the the
most likely candidates for non-simply connected cosmic topologies.

\smallskip

Cosmological constraints show that flat or nearly flat, very sightly positively curved, 
geometries are preferred over negatively curved ones. Combined with requirements of
homogeneity on the geometry, this selects as the most likely candidates the
flat tori and quotients (Bieberbach manifolds) or the sphere and quotient spherical
forms. We compare here the behavior of two among the most promising
spherical candidates, the quaternionic space and the Poincar\'e
homology sphere or dodecahedral space, and we show that, in the gravity
model based on the spectral action functional, they both behave like the
3-sphere in terms of the resulting model of inflation with slow-roll potential.
We then analyze the case of flat tori, and we show that these instead show
a distinctly different behavior in terms of possible models of inflation. 

\smallskip

Finally, in an appendix we discuss the case of lens spaces, where a
discrepancy in the existing mathematical literature on the Dirac spectrum
gives rise to a ``false positive" in terms of the relation between cosmic
topology and inflation. 

\smallskip

Our method consists of extending the non-perturbative calculation of the spectral 
action of the sphere given in \cite{uncanny} to all these other cases, by subdividing 
the Dirac spectrum into a union of arithmetic progressions with multiplicities that can be 
interpolated via values of polynomials at points of the spectrum. Then one
can apply the Poisson summation formula to each of these progressions and
obtain a complete explicit computation of the spectral action non-pertubatively. 

\section{The problem of cosmic topology}

The problem of cosmic topology is the question of whether the spatial topology of
the universe can be constrained on the basis of available cosmological data, especially
coming from the cosmic microwave background (CMB).  A general
introduction to cosmic topology is given in \cite{LaLu}.  

\smallskip

It was known since the mid '90s that the CMB anisotropies may produce constraints on
the geometry of the universe \cite{KaSpeSu}. In fact, the constraints on the $\Omega_0$ 
parameter favor a spatial geometry that is either flat or nearly flat, slightly positively 
curved (see \cite{dBL}). However, even with the curvature severely constrained by
cosmological data, there are still different possible multiconnected topologies that
support a homogeneous metric with given nearly flat constant curvature. 
The curvature constraints thus suggest 
that spherical space forms $S^3/\Gamma$, flat tori $T^3$ and Bieberbach
manifolds $T^3/\Gamma$ are all good possible candidates for cosmic topologies \cite{UKE}. 
Since the first year of WMAP data \cite{Sper}, the problem of cosmic topology became
especially interesting for the main reason that a multiconnected topology may be able
to account for some of the anomalies observed in the CMB anisotropies. 
In fact, the WMAP data suggested possible violations of statistical isotropy in the angular 
correlation function of the temperature fluctuations. The main anomalies are the quadrupole
suppression, the small value of the two-point temperature correlation function at angles
above 60 degrees, and the anomalous alignment of the quadrupole and octupole \cite{Teg}. 
These anomalies could be an indication of the presence of interesting 
(non simply connected) cosmic topologies.  

\smallskip

As discussed for instance in \cite{SouHa}, there are at present three main approaches
to investigating the question of cosmic topology.
\begin{itemize}
\item A search for multiple imaging in the CMB sky would reveal the periodicities caused by the
matching of sides of a fundamental domain for a manifold that is a compact quotient of a
model geometry (flat or spherical). This type of search is knows also as ``circles in the sky method".
\item A non-trivial cosmic topology is expected to violate the statistical isotropy of the angular power spectrum of CMB anisotropies, that is, the rotational invariance of $n$-point correlations.
\item Different cosmic topologies may also be detectable through correlation patterns of the CMB
anisotropy field which may be detectable through the coefficients of the expansion of
the field in spherical harmonics.
\end{itemize}

At present there are no conclusive results that either prove or disprove the presence of
a non-simply connected spatial topology in the universe. Although encouraging initial
results \cite{LWRL} suggested that one of the most widely studied candidate for
cosmic topology, the Poincar\'e homology sphere or dodecahedral space, could account
for the missing large angle correlations of the two-point angular correlation function of 
the temperature spectrum of the CMB, attempts 
to account for the quadrupole-octupole alignment  in this topology have failed \cite{WeGu}. 
A ``circle in the skies" search based on the first year of WMAP data \cite{CSSK} also failed to 
identify any non-simply connected topologies.  

\smallskip

An explicit description of all the different 
candidate spherical spaces was given in \cite{GLLUW}, with an analysis of how they may
be detectable through ``crystallographic method" through the presence of spikes in the
pair separation histogram for three dimensional catalogs of cosmic objects. 

\smallskip

In addition to the three approaches mentioned above, there has been recently also
an analysis of the problem of cosmic topology from the point of view of {\em residual
gravity acceleration}, \cite{RouRo}. This predicts that, in a non-simply connected topology which
is a quotient of either the flat 3-dimensional space or the sphere by a discrete group
of isometries, a test particle of negligible mass that feels the gravitational influence of
a nearby massive object should also feel a gravitational effect from the translates of the
same massive objects in nearby fundamental domains of the group action. This gives
rise to a gravitational effect qualitatively similar to dark energy. Due to symmetries,
it is shown in \cite{RouRo} that this effect vanishes at first order, but has
nontrivial third order effects. In the particular case of the Poincar\'e homology sphere,
it vanishes also at third order and only has non-trivial fifth order effects. It is also shown
that, in cases like tori $T^3$ with three different translation lengths, the residual gravity
acceleration effect tends to pull the space back to its most symmetric form with three
equal translation lengths. Thus, cosmology dynamically prefers the most symmetric
forms for a given topology.

\smallskip

Recently, predictions of possible cosmic topologies have also been obtained
within brane-world scenarios in string theory \cite{McInnes}.

\smallskip

\subsection{Laplace spectrum and cosmic topology}

What is especially interesting from our point of view is the fact that the way possible 
non-simply connected topologies manifest themselves in the CMB is mostly through properties
of the Laplace spectrum of these manifolds.  

More precisely, in cosmology it is customary to express the temperature fluctuations of the 
CMB as a series in the spherical harmonics $Y_{\ell m}$, of the form
\begin{equation}\label{sphereharmonics}
\frac{\Delta T}{T} = \sum_{\ell=0}^\infty \sum_{m=-\ell}^\ell a_{\ell m} Y_{\ell m}.
\end{equation}
One then looks at the correlation for the parameters $a_{\ell m}$. In the case of $S^3$,
the off diagonal terms vanish, 
\begin{equation}\label{diagcorrT}
\langle a_{\ell m} a^*_{\ell' m'} \rangle = C_\ell \delta_{\ell \ell'}
\delta_{m m'}, 
\end{equation}
while the diagonal terms $C_\ell$ give the temperature anisotropy power 
spectrum. In the case of a spherical space form $S^3/\Gamma$, with $\Gamma$ a finite
group of isometries, it is well known that these correlation functions in general 
no longer have vanishing off-diagonal components. The information on the different topologies 
is therefore encoded in the eigenfunctions of the Laplacian, which replace the usual spherical harmonics of $S^3$ in the computation of these correlation functions.

\medskip

Thus, an explicit computation of the spectrum and eigenfunctions of the Laplacian on
the candidate 3-manifolds, as in \cite{LWUGL}, \cite{RULW}, can be used to produce
simulated CMB skies for the different candidate topologies, which are then compared
to the WMAP data for the observed CMB.

\medskip

Various statistical tests have been developed to compare different candidate topologies and
search for a best fit with observational data. In particular, in the case of the simplest spherical
geometries, a comparison based on Bayesian analysis was given recently in \cite{NiarJaffe},
where simulated CMB maps for these different topologies are also exhibited.
The work \cite{NiarJaffe} compares the cosmological predictions of suppression of power at 
low $\ell$ for five different spherical manifolds: the simply connected case $S^3$,
the quaternionic space, and the three exceptional geometries, octahedral, truncated cubic, and dodecahedral.


\medskip

\section{The spectral action and cosmic topology}

In this paper we follow a very different point of view on the problem of cosmic
topology. We work within a particular theoretical model of gravity coupled to
matter, which arises from the noncommutative geometry models of particle
physics developed in \cite{CoSM} and more recently \cite{CCM}. These models
are based on extending ordinary spacetime to a product by small extra dimensions,
which, unlike in string theory models, are not manifolds but noncommutative spaces.

\smallskip

Within these models, one has a natural choice of an action functional, which
is the {\em spectral action} of \cite{ChCo}. This is essentially an action functional
for gravity on the product space $X\times F$, with $X$ the ordinary $4$-dimensional
spacetime manifold and $F$ the fiber noncommutative space. The large energy
asymptotic expansion of the spectral action functional delivers a Lagrangian
with gravity terms including the usual Einstein--Hilbert action with a cosmological
term, and additional conformal gravity terms. Moreover, in the asymptotic expansion
one also finds a coupled matter Lagrangian, which depends on the choice of
the non-commutative space $F$. It is shown in \cite{CCM} that, for a suitable
choice of $F$, the resulting Lagrangian recovers the full Lagrangian of an extension
of the minimal Standard Model with right handed neutrinos and Majorana mass terms.
More recent work \cite{BroSu} shows that a modified choice of the space $F$ leads
to a further extension of the Standard Model that incorporates supersymmetric QCD.

\smallskip

In all these cases, the main feature of these noncommutative geometry models
is that one has a non-perturbative action functional on $X\times F$, whose 
asymptotic expansion delivers a Lagrangian for particle physics coupled to
gravitational terms. In other words, gravity on the noncommutative product space $X\times F$
manifests itself as gravity coupled to matter on the ordinary (commutative) spacetime
manifold  $X$.

\subsection{The spectral action functional}

The generalization of Riemannian geometry in the world of noncommutative geometry 
is provided by the notion of spectral triples. These are data 
$(\cA,\cH,D)$, with $\cA$ the algebra of functions on the (possibly noncommutative)
space, $\cH$ the Hilbert space of square integrable spinors, and $D$ the Dirac operator.
The information corresponding to the metric tensor is encoded in the Dirac operator.

\smallskip

The spectral action functional of \cite{ChCo} is defined as $\Tr (f(D/\Lambda))$,
where $\Lambda$ is the energy scale and $f$ is a test function, usually a 
smooth approximation of a cutoff function. 
This is regarded as a spectral formulation of gravity in noncommutative
geometry.  This action functional has an asymptotic expansion at high
energies 
\begin{equation}\label{SpAct}
\Tr (f(D/\Lambda)) \sim \sum_{k\in {\rm DimSp}} f_{k} \Lambda^k {\int\!\!\!\!\!\!-} |D|^{-k} + f(0) \zeta_D(0)+ o(1) ,
\end{equation}
where $f_k= \int_0^\infty f(v) v^{k-1} dv$, $f_0=f(0)$, $f_{-2k}=(-1)^k \frac{k!}{(2k)!} f^{(2k)}(0)$, 
and integration is given by residues of zeta function $\zeta_D (s) = \Tr (|D|^{-s})$
at the points $k$ in the Dimension Spectrum, that is, at poles of the zeta function.

\smallskip

It suffices for our purposes to concentrate
only on the gravitational sector of the noncommutative
geometry model, because that is where we expect to see a
signature of cosmic topology to appear. This
means that, instead of computing the spectral action on
the product space $X\times F$, with a noncommutative
space $F$  that accounts for the matter terms in the 
Lagrangian arising from the asymptotic expansion,
we only compute it on the underlying commutative 
spacetime manifold $X$. 

More precisely, in Section \ref{S3recallSec} we recall the 
computation for the sphere case done in \cite{uncanny}. 
In the following sections we obtain explicit non-perturbative 
computations of the spectral actions on various 3-manifolds 
that are interesting candidates for cosmic topologies.

We then show in \S \ref{slowrollSec} that perturbations of the 
Dirac operator of the form $D^2\mapsto D^2 + \phi^2$, as 
considered in \cite{uncanny} for the sphere case, provide a 
slow-roll potential for a field conformally coupled to gravity,
which provides a mechanism for inflation. 

We then compute the spectral action and the
corresponding slow-roll parameters explicitly for the various
different topologies, two spherical ones and a flat case,
and we show that in the spherical cases the  
slow-roll parameters agree with those for the simply connected
topology while in the flat case they are different. 
We discuss separately the case of lens spaces in the appendix.

Since the spectral action is computed non-perturbatively,
the results are not confined to the very early universe near
unification energy, but extend to lower energies, so that
predictions about slow-roll parameters and cosmological
properties like tensor-to-scalar ratio and spectral index
can, in principle, be compared with observational data.
To obtain a more precise model that can be directly
compared with data one should also include further corrections
to the slow-roll parameters coming from the additional matter sector
(the noncommutative space $F$), which will
not be considered in this paper.

The main conclusion is that, in models of gravity coupled to
matter based on noncommutative geometry and the spectral
action functional,  there is a coupling of topology and inflation:
different spatial topologies can have a 
measurable effects on the tensor-to-scalar ratio and spectral index,
through their effects on the slow-roll parameters of a slow-roll
potential coming from perturbations of the Dirac operator and
non-perturbative effects in the spectral action functional.

\bigskip

\section{Recalling the case of $S^3$}\label{S3recallSec}

In this section we recall briefly the results of Chamseddine--Connes \cite{uncanny} on
the non-perturbative calculation of the spectral action for the 3-sphere, which we need later,
when we compare their result to the analogous computation in the cases 
of the other candidate cosmic topologies.

\medskip

\subsection{Euclidean model}

We are interested in investigating cosmological signatures of 
the {\em spatial} topology of
the universe. This means the topology of a spatial 3-dimensional 
section of the 4-dimensional Lorentzian spacetime describing 
the universe. It is customary, in working with the spectral action 
functional of noncommutative geometry, to Wick rotate to a
Euclidean model of gravity on a compact manifold. Thus, instead 
of working with a non-compact Lorentzian 4-manifold which is
topologically a cylinder $S\times \R$, with $S$ a compact 3-dimensional
Riemannian manifold, we Wick rotate and compactify to a Riemannian
4-manifold $X=S \times S^1$, where the size of $S$ is unaltered and
the size of the compactified $S^1$ acquires the meaning of a thermodynamic
parameter, an inverse temperature $\beta$, as in \cite{uncanny}.

The spectral action of a 3-dimensional manifold $S$ is related in \cite{uncanny} 
to the 4-dimensional geometry $X=S \times S^1$ by showing that the transform 
\begin{equation}\label{34transform}
k(x) = \int_x^\infty (u-x)^{-1/2} h(u)\, du
\end{equation}
relates the spectral action functionals for
the 3-dimensional and 4-dimensional geometries by
\begin{equation}\label{34asympt}
\Tr(h(D_X^2/\Lambda^2)) \sim 2 \beta \Lambda\, \Tr(k(D_S^2 /\Lambda^2)),
\end{equation}
with $\beta$ the size of the circle $S^1$ and the Dirac operator $D_X$ of
the form
\begin{equation}\label{DXop}
D_X =\left(\begin{array}{cc}  0 & D_S\otimes  1 + i \otimes D_{S^1} \\
D_S \otimes 1 - i \otimes D_{S^1} & 0     \end{array}\right),
\end{equation}
where $D_{S^1}$ has spectrum $\beta^{-1} (\Z + 1/2)$.

\medskip

\subsection{The Poisson summation formula}

The Poisson summation formula states that, for a test function $h$ in Schwartz space
$h \in \cS(\R)$, one has
\begin{equation}\label{Poisson}
\sum_{n\in \Z} h(n) = \sum_{n \in \Z} \widehat h(n),
\end{equation}
or the more general form
\begin{equation}\label{Poisson2}
\sum_{n\in \Z} h(x+\lambda n) =\frac{1}{\lambda} \sum_{n \in \Z} e^{\frac{2\pi i n x}{\lambda}} 
\widehat h(\frac{n}{\lambda}),
\end{equation}
with $\lambda\in \R^*_+$ and $x\in \R$, 
where $\widehat h$ is the Fourier transform
\begin{equation}\label{Fourier}
\widehat h(x) =\int_{\R} h(u)\, e^{-2\pi i ux}\, du.
\end{equation}

In \cite{uncanny} the Poisson summation formula is applied
to a test function of the form $h(u)=P(u) f(u/\Lambda)$, where $P(u)$
is a polynomial function that gives a smooth interpolation for the
multiplicites of the Dirac eigenvalues on the 3-sphere $S^3$
and $f$ is a smooth approximation to a cutoff function, used
in the spectral action functional. 
This allows for an explicit nonperturbative computation of the
spectral action functional in the case of the 3-sphere.

\medskip

As shown in \S 2.2 of \cite{uncanny},
for a sphere $S^3$ with radius $a$ the Dirac spectrum is
given by $\pm a^{-1} (\frac{1}{2}+n)$ for $n\in \Z$, with multiplicity
$n(n+1)$. The Poisson summation formula
as above gives a spectral  action of the form
\begin{equation}\label{S3spact}
\begin{array}{ll}
\Tr( f(D/\Lambda)) & =  (\Lambda a)^3 \widehat{f}^{(2)} (0) - \frac{1}{4} (\Lambda a) \widehat{f}(0) +O((\Lambda a)^{-k}) \\[3mm]
& = (\Lambda a)^3 \int_\R v^2 f(v)\, dv - \frac{1}{4} (\Lambda a) \int_\R f(v)\, dv +
O((\Lambda a)^{-k}) ,
\end{array}
\end{equation}
where $\widehat f^{(2)}$ denotes the Fourier transform of $v^2 f(v)$.

\medskip

\subsection{The spectral action in 4-dimensions} 

The corresponding computation of the spectral action for $S^3\times S^1$ is
done in \cite{uncanny} using Poisson summation 
\begin{equation}\label{PoissonS3S1}
 \sum_{(n,m)\in \Z^2} g(n+\frac{1}{2},m+\frac{1}{2}) = \sum_{(n,m)\in \Z^2} (-1)^{n+m} \widehat g(n,m),
\end{equation} 
for a function of two variables
\begin{equation}\label{guv}
 g(u,v) = 2 P(u) \, h(u^2 (\Lambda a)^{-2}+ v^2 (\Lambda \beta)^{-2}), 
\end{equation} 
where $P(u)$ is the polynomial that interpolates the multiplicities of
the spectrum on the sphere $S^3$ of radius $a$ and $h(D^2/\Lambda^2)$ is
the Schwartz function in the spectral action, and $\beta$ is the size of the circle $S^1$,
which has Dirac spectrum $\beta^{-1}(\Z +1/2)$.

One obtains then the spectral action on $S^3\times S^1$ using Poisson 
summation on $\Z^2$. This gives
\begin{equation}\label{FourierS3S1}
 \Tr(h(D^2/\Lambda^2)) = \widehat g(0,0) + O(\Lambda^{-k}) 
\end{equation}
for any $k>0$, where
\begin{equation}\label{gFourier}
 \widehat g(n,m) =\int_{\R^2} g(u,v) e^{-2\pi i(xu+yv)}\, du\, dv, 
\end{equation} 
and the error term $\sum_{(n,m)\neq (0,0)} (-1)^{n+m} \widehat g(n,m)$
is estimated to be smaller than $\Lambda^{-k}$.  One obtains from \eqref{FourierS3S1}
\begin{equation}\label{SpActS3S1}
\Tr(h(D^2/\Lambda^2)) = \pi \Lambda^4 a^3 \beta \int_0^\infty u \,h(u)\, du - \frac{1}{2} \pi \Lambda a \beta
\int_0^\infty h(u) \, du + O(\Lambda^{-k}) .
\end{equation}

One can consider particular classes of
test functions $h$ which approximate well enough an even cutoff function
on the Dirac spectrum. The class of functions used in \cite{uncanny}
is test functions of the form $h(x)=P(\pi x) e^{-\pi x}$ with $P$ a
polynomial, and in particular, among these, a good approximation to
a cutoff function  given by the test functions $h_n(x^2)$ with
\begin{equation}\label{hn}
 h_n(x) = \sum_{k=0}^n \frac{(\pi x)^k}{k!} e^{-\pi x}. 
\end{equation}

\medskip

\section{Slow-roll potential from the spectral action}\label{slowrollSec}

We show here that the perturbations of the Dirac operator
considered in \cite{uncanny}, of the form $D^2 \mapsto D^2 
+\phi^2$, give rise to a slow-roll potential for
inflation. We compute the corresponding slow-roll 
parameters in the case of $S^3$. We then compute
the potential and slow-roll parameters in the case of
the other candidate cosmic topologies and we compare
them with the case of $S^3$.

\subsection{Nonperturbative corrections and slow-roll potential}

One of the most interesting aspects of the results of \cite{uncanny} is that,
under the perturbation $D^2 \mapsto D^2 
+\phi^2$ of the Dirac operator, one finds a potential
$V(\phi)$ for a scalar field $\phi$ conformally coupled to gravity,
which at low energies behaves like a quartic Higgs potential, but
which has additional nonperturbative corrections, which have the
effect, at higher energies of flattening out the form of the potential
so that it is asymptotic to a constant. This gives it
the typical form of the slow-roll potentials used in
models of inflation in cosmology.

\medskip

The replacement $D^2 \mapsto D^2 + \phi^2$, corresponding
to a shift $h(u) \mapsto h(u+ \phi^2/\Lambda^2)$ in the test function 
(assumed of the form $h(x)=P(\pi x) e^{-\pi x}$ as above) produces a potential
for the field $\phi$, which, for sufficiently small values of the parameter 
$x=\phi^2/\Lambda^2$, recovers the usual quartic potential for the field 
$\phi$, conformally coupled to gravity, which on $S^3 \times S^1$ is of the form
\begin{equation}\label{HpotentialS3}
- \pi \Lambda^2 \beta a^3 \int_0^\infty h(v) dv \,\, \phi^2 + \frac{1}{2} \pi \beta a h(0) \,\, \phi^2
+ \frac{1}{2} \pi \beta a^3 h(0)\,\, \phi^4 .
\end{equation}
These correspond to a term of the form $\int_X R\, \phi^2 \, dvol$, giving the
conformal coupling to gravity, from the 4th Seeley-de Witt coefficient, together
with a quadratic mass term and a quartic potential , 
respectively from the second and 4th coefficient.

\medskip
 
However, for larger values of the parameter $x=\phi^2/\Lambda^2$, the
potential obtained from the nonperturbative calculation of \cite{uncanny}
levels out. Theorem 7 of \cite{uncanny} shows that one has on $S^3\times S^1$
\begin{equation}\label{SpActHiggsS3}
\begin{array}{c}
\Tr(h((D^2 + \phi^2)/\Lambda^2))) = 2\pi \Lambda^4 \beta a^3 \int_0^\infty h(\rho^2)\rho^3 d\rho
- \pi \Lambda^2 \beta a \int_0^\infty h(\rho^2) \rho d\rho \\[3mm]
+ \pi \Lambda^4 \beta a^3\, \cV(\phi^2/\Lambda^2) +\frac{\pi}{2} \Lambda^2 \beta a\, \cW(\phi^2/\Lambda^2)
 +\epsilon(\Lambda),
 \end{array}
\end{equation}
where the error term $\epsilon(\Lambda)$ is exponentially small in $\Lambda$
and the funcions $\cV$ and $\cW$ are given by
\begin{equation}\label{VW}
\cV(x)= \int_0^\infty u (h(u+x)-h(u))du, \ \ \ \ \ \cW(x)= \int_0^x h(u)du. 
\end{equation}

\medskip

This is the typical behavior expected from a slow-roll potential used in 
scenarios for inflationary cosmology based on the Standard Model of
elementary particles, as in the recent paper \cite{dSHW}. In particular,
we are interested in deriving the associated slow-roll parameters.

\medskip

\subsection{Slow-roll parameters}

A way to obtain models of inflation with slow roll potential is to
have a theory with a non-minimal coupling of a scalar field to
gravity via the curvature $R$. For a version of a Higgs based
inflation see \cite{dSHW}. We show here that the nonperturbative
corrections to the Higgs potential in the spectral action obtained
in \cite{uncanny} present a similar scenario.

\smallskip

Notice that, in the derivation of the slow-roll potential from the
spectral action, we have replaced the Minkowskian spacetime geometry
with a compactified Euclidean model in order to compute the spectral
action nonperturbatively and then derive the slow-roll potential from
the perturbation of the Dirac operator. However, once we have
obtained a Lagrangian for gravity coupled to a scalar field $\phi$
that will be responsible for inflation, we can continue the same
Lagrangian back to Minkowskian signature and consider the effects
of the slow-roll potential over a Minkowskian spacetime given by a Friedmann 
metric with assigned topology on the spatial sections.

\medskip

Consider a Minkowskian space-time metric of the form
\begin{equation}\label{Friedmannds}
 ds^2 = a(t)^2 ds^2_S - dt^2, 
\end{equation}
where $ds_S^2$ is the assigned Riemannian homogeneous
metric on the 3-manifold $S$ (the candidate cosmic topology)
and $a(t)$ is the scale factor.
 
\smallskip

In models of inflations based on a scalar field with a slow-roll
potential $V(\phi)$, the accelerated expansion phase $\ddot{a}/a >0$
is governed by the equation
$$ \frac{\ddot{a}}{a} = H^2 (1-\epsilon), $$
where the Hubble parameter $H^2(\phi)$ is related to the slow roll potential $V(\phi)$ by
$$ H^2(\phi) \left( 1-\frac{1}{3} \epsilon(\phi) \right) = \frac{8\pi}{3 m_{Pl}^2} \, V(\phi), $$
where $m_{Pl}$ is the Planck mass and $\epsilon(\phi)$ is the first slow-roll 
parameter satisfying the equation of state
\begin{equation}\label{epsilonphi}
\epsilon(\phi)= \frac{m_{Pl}^2}{16\pi} \left( \frac{V^\prime(\phi)}{V(\phi)} \right)^2. 
\end{equation}
The inflationary phase is characterized by $\epsilon(\phi)<1$. The second
slow-roll parameter has the form
\begin{equation}\label{etaphi}
\eta(\phi) = \frac{m_{Pl}^2}{8\pi}  
 \left( \frac{V^{\prime\prime}(\phi)}{V(\phi)} \right)
- \frac{m_{Pl}^2}{16\pi} \left( \frac{V^\prime(\phi)}{V(\phi)} \right)^2.
\end{equation}

These parameters enter in two important measurable quantities: 
the spectral index and the tensor-to-scalar ratio, which are given
respectively by
\begin{equation}\label{nsrparam}
\begin{array}{ll}
n_s = & 1 - 6 \epsilon + 2 \eta \\
r = & 16 \epsilon.
\end{array}
\end{equation}

\medskip

We remark that, from the cosmological viewpoint, the model we
consider here will only be a toy model, in the sense that, as in
\cite{uncanny} we only look at the purely gravitational part of the
spectral action, and we do not consider the effect of the presence
of matter coming from the presence of the additional noncommutative
space as extra dimensions. This simplification has the advantage that
it allows us to focus only on the nonperturbative effects on the Higgs
potential, without having to carry around additional terms that are
not directly affected by the 3-dimensional spatial topology. However, 
one should keep in mind that the resulting slow-roll parameters will
also be affected by the matter contributions, as described in \S 4 of
\cite{MaPie}. So, in particular, the values we obtain here for these
parameters, in the simplifying assumption that drops the matter
part, need not meet the observational constraints. The main
point for us is to show that there is a contribution to these slow-roll
parameters that can be {\em different} from that of the sphere in 
certain candidate cosmic topologies such as the flat tori or
equal to that of the sphere in other candidate topologies such
as quaternionic or dodecahedral spaces. For this reason we drop 
the matter terms that would not differ in the various cases.

\medskip

\subsection{Slow-roll parameters for the $S^3$-topology}

We now compute the slow-roll parameters resulting from the
nonperturbative corrections to the Higgs potential of \cite{uncanny},
in the case where the underlying spatial topology is the
3-sphere.

\smallskip

\begin{thm}\label{slowrollS3}
The slow-roll potential
\begin{equation}\label{S3Vphi}
 V(x) = \pi\Lambda^4 \beta a^3 \cV(\phi^2/\Lambda^2) + \frac{\pi}{2} \Lambda^2 \beta a \cW(\phi^2/\Lambda^2), 
\end{equation} 
with $\cV$ and $\cW$ as in \eqref{VW}, and $x=\phi^2/\Lambda^2$, has slow-roll parameters
\begin{equation}\label{epsilonS3}
\epsilon(x)= \frac{m_{Pl}^2}{16\pi} \left( \frac{h(x) - 2 (\Lambda a)^2 \int_x^\infty h(u)du }
{\int_0^x h(u)du + 2 (\Lambda a)^2 \int_0^\infty u (h(u+x)-h(u)) du}\right)^2
\end{equation}
and
\begin{equation}\label{etaS3}
\begin{array}{rl}
\eta(x) =& \displaystyle{\frac{m_{Pl}^2}{8\pi} \frac{h^\prime (x)+2 (\Lambda a)^2 h(x)}{\int_0^x h(u)du + 2 (\Lambda a)^2 \int_0^\infty u (h(u+x)-h(u)) du}} \\[4mm] - & \displaystyle{ \frac{m_{Pl}^2}{16\pi} \left( \frac{h(x) - 2 (\Lambda a)^2 \int_x^\infty h(u)du }
{\int_0^x h(u)du + 2 (\Lambda a)^2 \int_0^\infty u (h(u+x)-h(u)) du}\right)^2} , \end{array}
\end{equation}
written in the variable $x=\phi^2/\Lambda^2$.
\end{thm}

\proof We have, as in Lemma 8 of \cite{uncanny}, 
$$ \cV'(x) = - \int_x^\infty h(u)du \ \ \ \text{ and } \ \ \  \cV^{\prime\prime}(x) =h(x), $$
while $\cW'(x) = h(x)$ and $\cW^{\prime\prime}(x)=h^\prime(x)$. So, if we write
$$ A= \frac{1}{2} \left( \frac{V^\prime(\phi)}{V(\phi)} \right)^2 \ \ \ \text{ and } \ \ \ 
 B=  \left( \frac{V^{\prime\prime}(\phi)}{V(\phi)} \right), $$
so that 
$$ \epsilon = \frac{m_{Pl}^2}{8\pi} A \ \ \  \text{ and } \ \ \  \eta = \frac{m_{Pl}^2}{8\pi} (B -A), $$
we find
$$ A=  \frac{1}{2} \left( \frac{h(x) - 2 (\Lambda a)^2 \int_x^\infty h(u)du }
{\int_0^x h(u)du + 2 (\Lambda a)^2 \int_0^\infty u (h(u+x)-h(u)) du}\right)^2 $$
and
$$ B =  \frac{h^\prime (x)+2 (\Lambda a)^2 h(x)}{\int_0^x h(u)du + 2 (\Lambda a)^2 \int_0^\infty u (h(u+x)-h(u)) du}. $$
\endproof

\begin{rem}\label{noscale} {\rm The slow-roll parameters obtained in this way are independent of
the scale $\beta$, as one should expect since that was an artifact introduced by
our passing to a Euclidean model to perform calculations with the spectral action,
while they depend on both the energy scale $\Lambda$ and the scale factor $a$,
but only through their product $\Lambda a$.  This again fits in well with
cosmological models, since we know that, for a cosmology described by a Friedmann
metric \eqref{Friedmannds}, the time dependence of the energy
scale factor is related through $\Lambda(t)\sim 1/a(t)$, so that their product is
a constant $C$ independent of time. }
\end{rem}

Thus, we can rewrite \eqref{epsilonS3} and \eqref{etaS3} for the 3-sphere
in the form
\begin{equation}\label{epsilonetaS3}
\begin{array}{rl}
\epsilon(x)= & \displaystyle{\frac{m_{Pl}^2}{16\pi} \left( \frac{h(x) - 2 C \int_x^\infty h(u)du }
{\int_0^x h(u)du + 2 C \int_0^\infty u (h(u+x)-h(u)) du}\right)^2 } \\[4mm]
\eta(x) =& \displaystyle{\frac{m_{Pl}^2}{8\pi} \,\, \frac{h^\prime (x)+2\pi C\, h(x)}{\int_0^x h(u)du + 2  C\int_0^\infty u (h(u+x)-h(u)) du}} \\[4mm] - & \displaystyle{ \frac{m_{Pl}^2}{16\pi} \left( \frac{h(x) - 2 C \int_x^\infty h(u)du }{\int_0^x h(u)du + 2\pi C  \int_0^\infty u (h(u+x)-h(u)) du}\right)^2} .
\end{array}
\end{equation}
 
We now compare this inflation model derived from the nonperturbative spectral action
on the sphere with the case of other nontrivial topologies.

\bigskip

\section{The quaternionic cosmology and the spectral action}

Let $Q8$ denote the group of quaternion units 
$\{ \pm 1, \pm i, \pm j, \pm k\}$. It acts on the 3-sphere, with
the latter identified with the group $SU(2)$. The resulting
quotient manifold $SU(2)/Q8$ plays an interesting role as a possible
cosmic topology candidate, in view of the recent results of \cite{NiarJaffe}
on the statistical comparison of various spherical space forms in
terms of the best fit for either the power spectrum $C_\ell$ or the
off-diagonal part of the correlation matrices.

As shown in the study of correlation matrices, as exhibited in Figure 3 of \cite{NiarJaffe},
the quaternionic space, unlike the other nontrivial topologies considered in their study,
shows no additional structure in the off-diagonal correlations with respect to the spherical
case. Thus, the analysis of off-diagonal terms in the correlation functions does not suppress
the Bayesian factor of the quaternionic space, while it suppresses those of all the other
nontrivial topologies. While the other model comparison carried out in \cite{NiarJaffe}
using the power spectrum $C_\ell$ does not favor this topology, the particular behavior
of the off-diagonal terms seems sufficiently interesting to develop additional possible
tests for comparing the quaternionic geometry $SU(2)/Q8$ to the ordinary spherical
geometry.

\medskip

\subsection{The Dirac spectra for $SU(2)/Q8$}

As we show here, the main reason why the case of $SU(2)/Q8$ can be
treated with the same technique used in \cite{uncanny} for the sphere
$S^3$ is because the Dirac spectrum is given in terms of arithmetic
progressions indexed over the integers, so that one can again apply the
same type of Poisson summation formula. This is not immediately the
case for other spherical geometries.

More precisely, we recall from \cite{Gin} that one can endow the
3-manifold $SU(2)/Q8$ with a 3-parameter family of homogeneous
metrics, depending on the parameters $a_i\in \R^*$, $i=1,2,3$. The
different possible spin structures $\epsilon_j$ on $SU(2)/Q8$ correspond to the
four group homomorphisms $Q8 \to \Z/2\Z$ with $\epsilon_0\equiv 1$ and
$\Ker(\epsilon_j)=\{ \pm 1, \pm \sigma_j \}$, with $\sigma_j$ the Pauli matrices.
The Dirac operator for each of these spin structures and its spectrum are
computed explicitly in \cite{Gin}. The case we are interested in here is 
only the one where the metric has parameters $a_1=a_2=a_3=1$,
for which $SU(2)/Q8$ is a spherical space form. For this case the Dirac
spectrum was also computed in \cite{Bar}. 

In this case, see Corollary 3.2 of \cite{Gin}, the Dirac spectrum for
$SU(2)/Q8$ with the spherical metric $a_1=a_2=a_3=1$, is given
in the case of the spin structure $\epsilon_0$ by 
\begin{equation}\label{SpectrumN1}
\left\{ \begin{array}{lll}
\frac{3}{2} + 4k & \text{with multiplicity} & 2(k+1)(2k+1) \\[2mm]
\frac{3}{2} + 4k +2 & \text{with multiplicity} & 4k(k+1) \\[2mm]
- \frac{3}{2} - 4k -1 & \text{with multiplicity} & 2k(2k+1)  \\[2mm]
- \frac{3}{2} - 4k -3 & \text{with multiplicity} & 4(k+1)(k+2),
\end{array}\right.
\end{equation}
where $k$ runs over $\N$.  For all the other
three spin structures $\epsilon_j$, $j=1,2,3$, the spectrum is given by
\begin{equation}\label{SpectrumN2}
\left\{ \begin{array}{lll}
\frac{3}{2} + 4k & \text{with multiplicity} & 2k(2k+1) \\[2mm]
\frac{3}{2} + 4k +2 & \text{with multiplicity} & 4(k+1)^2 \\[2mm]
- \frac{3}{2} - 4k -1 & \text{with multiplicity} & 2(k+1)(2k+1)  \\[2mm]
- \frac{3}{2} - 4k -3 & \text{with multiplicity} & 4(k+1)^2,
\end{array}\right.
\end{equation}
again with $k\in \N$.

\subsection{Trivial spin structure: nonperturbative spectral action}

By replacing $k$ with $-k-1$ in the third row and $k$ with $-k-2$ in the fourth row,
we rewrite the spectrum \eqref{SpectrumN1} in the form
\begin{equation}\label{SpectrumN1bis}
\left\{ \begin{array}{lll}
\frac{3}{2} + 4k & \text{with multiplicity} & 2(k+1)(2k+1) \\[2mm]
\frac{3}{2} + 4k +2 & \text{with multiplicity} & 4k(k+1), 
\end{array}\right.
\end{equation}
where now $k$ runs over the integers $\Z$. This expresses the spectrum in
terms of two arithmetic progressions indexed over the integers. Now the condition
that allows us to apply the Poisson summation formula as in \cite{uncanny}
is the fact that the multiplicities can be expressed in terms of a smooth function of $k$.
This is the case, since the multiplicites in \eqref{SpectrumN1bis} for an eigenvalue $\lambda$
are given, respectively, by the functions $P_1(\lambda)$ and $P_2(\lambda)$ with
\begin{equation}\label{P12}
\begin{array}{ll}
P_1(u) & =\displaystyle{ \frac{1}{4} u^2 + \frac{3}{4} u + \frac{5}{16} } \\[5mm]
P_2(u) & = \displaystyle{ \frac{1}{4} u^2 - \frac{3}{4} u - \frac{7}{16} }.
\end{array}
\end{equation}

We then obtain an explicit nonperturbative calculation of the spectral action for
$SU(2)/Q8$ as follows.

\begin{thm}\label{SpAcQ8thm}
The spectral action on the 3-manifold $S=SU(2)/Q8$, with the trivial spin structure, is given by
\begin{equation}\label{SpAcQ80}
\Tr(f(D/\Lambda))=\frac{1}{8} (\Lambda a)^3 \widehat f^{(2)}(0) - \frac{1}{32} (\Lambda a) \widehat f(0) + \epsilon(\Lambda),
\end{equation}
with $a$ the radius of the 3-sphere $SU(2)=S^3$, 
with the error term satisfying $|\epsilon(\Lambda)|=O(\Lambda^{-k})$ for all $k>0$,
and with $\widehat f^{(k)}$ denoting the Fourier transform of $v^k f(v)$ as above.
Namely, the spectral action for $SU(2)/Q8$ is $1/8$ of the spectral action for $S^3$.
\end{thm}

\proof Consider a test function for the Poisson summation formula which is
of the form 
$$ h(u)=g(4u + \frac{s}{2}), \ \ \  \text{ for some } \ \  s\in \Z. $$
Then \eqref{Poisson} gives
\begin{equation}\label{g4nPoisson}
 \sum_{n\in \Z} g(4n + \frac{s}{2}) = \sum_{n\in \Z}  \frac{1}{4} 
\exp( \frac{ i\pi s n }{4} ) \, \widehat g(\frac{n}{4}), 
\end{equation}
which we apply to $g_i(u) = P_i(u) f(u/\Lambda)$, with $P_i$ as in \eqref{P12} and $f$ the
Schwartz function in the spectral action approximating a cutoff.

This gives an expression for the spectral action on $S=SU(2)/Q8$ with the trivial
spin structure, and with the sphere $S^3=SU(2)$ of radius one, which is of the form
\begin{equation}\label{SpAcSU2Q80g}
\begin{array}{rl}
\Tr(f(D/\Lambda)) = & \displaystyle{ \sum_\Z g_1(4n + \frac{3}{2}) + \sum_\Z  g_2(4n + \frac{7}{2}) } 
\\[4mm]
= & \displaystyle{ \sum_\Z \frac{1}{4} \exp(\frac{3\pi i n}{4}) \widehat g_1(\frac{n}{4}) + \sum_\Z \frac{1}{4}
\exp(\frac{7\pi i n}{4}) \widehat g_2(\frac{n}{4}). }
\end{array}
\end{equation}

Assuming that $f$ is a Schwartz function, then $g_i$ is also Schwartz, hence so is $\widehat g_i$. Therefore, for each $k\in \N$, we get an estimate of the form
$$ \sum_{n\neq 0} \frac{1}{4} | \widehat g_i(\frac{n}{4}) | \leq C_k \Lambda^{-k}. $$
This shows that we can write the right hand side of \eqref{SpAcSU2Q80g} as the
terms involving $\widehat g_i(0)$ plus an error term that is of order $O(\Lambda^{-k})$.

One then computes
\begin{equation}\label{g10Q8}
\widehat g_1(0) = \frac{1}{4} \Lambda^3 \widehat f^{(2)}(0) + \frac{3}{4} \Lambda^2 \widehat f^{(1)}(0) +
\frac{5}{16} \Lambda \widehat f(0).
\end{equation}
Similarly, on has
\begin{equation}\label{g20Q8}
\widehat g_2(0) = \frac{1}{4} \Lambda^3 \widehat f^{(2)}(0) - \frac{3}{4} \Lambda^2 \widehat f^{(1)}(0) -
\frac{7}{16} \Lambda \widehat f(0), 
\end{equation}
so that one obtains for the spectral action in \eqref{SpAcSU2Q80g}
\begin{equation}\label{SpAcQ80add}
\begin{array}{rl}
\Tr(f(D/\Lambda))=& \frac{1}{4} \left( \widehat g_1(0) + \widehat g_2(0) \right)  + O(\Lambda^{-k}) \\[4mm]
= & \frac{1}{8} \Lambda^3 \widehat f^{(2)}(0) - \frac{1}{32} \Lambda \widehat f(0) + O(\Lambda^{-k}).
\end{array}
\end{equation}
The case with the 3-sphere $SU(2)=S^3$ of radius $a$ is then analogous, with the
spectrum scaled by a factor of $a^{-1}$, which is like changing $\Lambda$ to $\Lambda a$
in the expressions above, so that one obtains \eqref{SpAcQ80}.
\endproof

\subsection{Nontrivial spin structures: nonperturbative spectral action}

The computation of the spectral action on $SU(2)/Q8$ in the case of the
non-trivial spin structures $\epsilon_j$ with $j=1,2,3$ is analogous. One
starts with the Dirac spectrum \eqref{SpectrumN2} and writes it in the form
of two arithmetic progressions indexed over the integers
\begin{equation}\label{SpectrumN2bis}
\left\{ \begin{array}{lll}
\frac{3}{2} + 4k & \text{with multiplicity} & 2k(2k+1) \\[2mm]
\frac{3}{2} + 4k +2 & \text{with multiplicity} & 4(k+1)^2. 
\end{array}\right.
\end{equation}
In this case one again has polynomials interpolating the values of the multiplicities. 
They are of the form
\begin{equation}\label{P12N2}
\begin{array}{ll}
P_1(u) & =\displaystyle{ \frac{1}{4} u^2 - \frac{1}{4} u - \frac{3}{16} } \\[5mm]
P_2(u) & = \displaystyle{ \frac{1}{4} u^2 + \frac{1}{4} u + \frac{1}{16} }.
\end{array}
\end{equation}
We then obtain the following result.

\begin{thm}\label{SpAcQ8thm}
The spectral action on the 3-manifold $S=SU(2)/Q8$, for any of the non-trivial spin structures
$\epsilon_j$, $j=1,2,3$, is given by the same expression \eqref{SpAcQ80} as in the case of
the trivial spin structure $\epsilon_0$.
\end{thm}

\proof It is enough to observe that the sum of the two polynomials \eqref{P12N2} that
interpolate the spectral multiplicities,
$$ P_1(u)+P_2(u)  = \frac{1}{2} u^2 - \frac{1}{8} $$
is the same as in the case \eqref{P12} of the trivial spin structure. One then has
the same value of
$$ \frac{1}{4} \widehat g_1(0) + \frac{1}{4} \widehat g_2(0) =\frac{1}{4} \int_\R (P_1(u)+P_2(u))\, f(u/\Lambda)\, du, $$
which gives the spectral action up to an error term of the order of $O(\Lambda^{-k})$.
\endproof

\medskip

\subsection{Slow-roll potential and parameters for the quaternionic space}

We compute now the slow-roll potential and slow-roll parameters for the
case of the quaternionic cosmic topology $S=SU(2)/Q8$. We first compute the
spectral action in the Euclidean 4-dimensional model $S\times S^1$, from
which we obtain the slow-roll potential by a perturbation of the Dirac operator
as in the case of $S^3$. 

\begin{thm}\label{SpAcQ8S1thm}
The spectral action on the 4-manifold $S\times S^1$ with $S=SU(2)/Q8$ is
given by
\begin{equation}\label{SpAcQ8S1}
\Tr(h(D^2/\Lambda^2)) =\frac{\pi}{8} \Lambda^4 a^3\beta \int_0^\infty u \, h(u)\, du -
\frac{\pi}{16} \Lambda^2 a\beta \int_0^\infty h(u)\, du + O(\Lambda^{-k}),
\end{equation}
namely $1/8$ of the spectral action for $S^3\times S^1$.
\end{thm}

\proof
The eigenvalues for the operator $D^2/\Lambda^2$ on $S\times S^1$ are 
$$ (4k + \frac{s}{2})^2 (\Lambda a)^{-2} + (m + \frac{1}{2})^2 (\Lambda \beta)^{-2}, $$
for $SU(2)=S^3$ of radius $a$ and $S^1$ of radius $\beta$, with multiplicities
$2 P_i(u)$, where $P_i(u)$ are the polynomials \eqref{P12} and \eqref{P12N2} 
that interpolate the spectral densities for $S=SU(2)/Q8$ and the integer $s$ also 
varies according to the arithmetic progressions in the spectrum \eqref{SpectrumN1bis} or \eqref{SpectrumN2bis}.

For a given integer $s$, the Poisson summation formula over $\Z^2$ gives
\begin{equation}\label{gnmQ8}
\sum_{(n,m)\in \Z^2} g(4n +\frac{s}{2}, m+\frac{1}{2}) =
\sum_{(n,m)\in \Z^2} \frac{1}{4} \exp(\frac{\pi i n s}{4})\, (-1)^m\, \widehat g (\frac{n}{4},m),
\end{equation}
where $g$ is a Schwartz function of the form \eqref{guv}. We have two functions
\begin{equation}\label{guvQ8}
 g_i (u,v) = 2 P_i(u)\, h(u^2 (\Lambda a)^{-2} + v^2 (\Lambda \beta)^{-2}),
\end{equation} 
with $P_i$ as in  \eqref{P12} and \eqref{P12N2}, respectively, for the trivial
and non-trivial spin structure.

In the case of the trivial spin structure one writes the spectral action as
\begin{equation}\label{SpAcQ8S1g}
\begin{array}{rl}
\Tr(h(D^2/\Lambda^2)) = & \displaystyle{ \sum_{\Z^2} g_1(4n+\frac{3}{2},m+\frac{1}{2})
+\sum_{\Z^2} g_2(4n +\frac{7}{2}, m+\frac{1}{2}) } \\[4mm]
=& \displaystyle{ \sum_{\Z^2} \frac{1}{4} \exp(\frac{3\pi i n}{4}) (-1)^m \widehat g_1(\frac{n}{4},m)} \\[4mm]
+& \displaystyle{\sum_{\Z^2} \frac{1}{4} \exp(\frac{7\pi i n}{4}) (-1)^m \widehat g_2(\frac{n}{4},m).}
\end{array}
\end{equation} 

The main term that contributes to \eqref{SpAcQ8S1g} is
\begin{equation}\label{g00Q8}
\begin{array}{l}
\widehat g_1(0,0) + \widehat g_2(0,0)  \\[4mm]
 \displaystyle{
= 2\Lambda^2 a\beta \int_{\R^2} \left( P_1(\Lambda ax)+
P_2(\Lambda ax)\right) \, h(x^2 + y^2)\, dx\,dy } \\[4mm]
= \displaystyle{ \Lambda^2 a\beta \int_{\R^2} \left( (\Lambda a)^2 x^2 - \frac{1}{4} \right)\,
h(x^2+y^2) \, dx\, dy } \\[4mm]
= \displaystyle{ \pi \Lambda^4 a^3 \beta \int_0^\infty h(\rho^2)\, \rho^3 \, d\rho
- \frac{\pi}{2} \Lambda^2 a \beta \int_0^\infty h(\rho^2) \, \rho\, d\rho.}
\end{array}
\end{equation}
The error term 
$$ \sum_{(n,m)\neq (0,0)}  \frac{1}{4} |\widehat g_i(\frac{n}{4},m)| $$
can be estimated as in \cite{uncanny} and is of the order of $O(\Lambda^{-k})$, for all $k>0$.

The result for the non-trivial spin structures is the same, since we have seen that
the sum $P_1(u)+P_2(u)$ is the same. This gives \eqref{SpAcQ8S1} after 
the change of variables 
$$ \int_0^\infty h(\rho^2)\, \rho^3 \, d\rho =\frac{1}{2} \int_0^\infty u\, h(u)\, du  \ \ \ \text{ and } \ \ \
\int_0^\infty h(\rho^2)\, \rho \, d\rho =\frac{1}{2} \int_0^\infty h(u)\, du. $$
\endproof

Since the spectral action for $S\times S^1$ in this case only differs from the
one of $S^3\times S^1$ by a the multiplicative factor $1/8$, we obtain the
following for the slow-roll potential and the slow-roll parameters.

\begin{prop}\label{slowrollQ8}
The slow-roll potential for $S=SU(2)/Q8$ is
$$ V(\phi) = \frac{1}{8} \pi\Lambda^4 \beta a^3 \cV(\phi^2/\Lambda^2) + \frac{1}{16}\pi \Lambda^2 \beta a \cW(\phi^2/\Lambda^2), $$
with $\cV$ and $\cW$ as in \eqref{VW}, and the slow-roll parameters are the same
\eqref{epsilonetaS3} as for the 3-sphere.
\end{prop}

Thus, in this case the slow-roll inflation model derived from the spectral action
does not distinguish the standard cosmic topology $S^3$ from the quaternionic
case $SU(2)/Q8$.

\bigskip

\section{Poincar\'e homology sphere: dodecahedral cosmology}

The Poincar\'e homology sphere, which is the quotient of the 3-sphere $S^3$
by the binary icosahedral group $\Gamma$, is also commonly referred to as the dodecahedral 
space, due to the fact that the action of $\Gamma$ on $S^3$ has a fundamental domain
that is a dodecahedron. The dodecahedral space is obtained by gluing together
opposite faces of a dodecahedron with the shortest clockwise twist that matches 
the faces. This space has been regarded as a likely candidate for the cosmic
topology problem and extensively studied for testable cosmological signatures
with all the methods presently available, \cite{CLLLRW}, \cite{GLLUW},
\cite{LaLu}, \cite{LWRL}, \cite{NiarJaffe}, \cite{RULW}, \cite{RouRo}, \cite{WeGu},
\cite{WLU}.

In particular, the three-year WMAP results confirmed the main anomalies: quadrupole
suppression, small value of the two-point temperature correlation function at large angles, 
and quadrupole--octupole alignment.  A recent analysis \cite{CLLLRW} 
of the Poincar\'e dodecahedral space based on the explicit computation of the
Laplace spectrum and the construction of the resulting simulated CMB sky with
more precise estimates of higher modes up to $\ell \sim 30$ finds a good match
to the WMAP data regarding the two-point temperature correlation function. 
Thus, the dodecahedral space remains at present one of
the most likely candidates, although it fails to account for other anomalies
like the quadrupole--octupole alignment \cite{WeGu}.

We give here the explicit computation of the spectral action functional for
the dodecahedral space, and we show then in \S \ref{slorolldodecaSec} 
that, in our model, from the point of view of the resulting inflation slow-roll parameters, 
the dodecahedral space behaves like the sphere, so that it cannot be ruled
out as a candidate cosmic topology in a gravity model based on the
spectral action.

\subsection{Generating functions for spectral multiplicities}

To compute explicitly the Dirac spectrum of the Poincar\'e homology sphere, we use 
a general result of B\"ar \cite{Bar}, which gives a formula for the generating function
of the spectral multiplicities of the Dirac spectrum on space forms of positive curvature.

In the generality of \cite{Bar}, one considers a manifold $M$ that is a quotient
$M=S^n/\Gamma$ of an $n$-dimensional sphere, $n\geq 2$, with the standard
metric of curvature one, with $\Gamma \subset SO(n+1)$ a finite group acting
without fixed points. It is shown in \cite{Bar} that 
the classical Dirac operator on $S^n$ has spectrum
\begin{equation}\label{SpecSn}
\pm \left( \frac{n}{2} + k \right), k \geq 0, \ \ \  \text{ with multiplicities } \ \ 2^{[n/2]}{k + n - 1 \choose k}.
\end{equation}
The eigenvalues of $M$ are the same as the eigenvalues of $S^n$, but with smaller multiplicities.  The spin structures of M are in 1-1 corrrespondence with homomorphisms  $\epsilon: \Gamma \rightarrow Spin(n+1)$, such that $\Theta \circ \epsilon = id_{\Gamma}$, where $\Theta$ is simply the double cover map from $Spin(n+1)$ to $SO(n+1)$. If $D$ is the Dirac operator on M, then to specify the spectrum of $M$, for one of these spin structures, one just needs to know the multiplicities, $m(\pm (n/2 + k))$, $k \geq 0$.  These are encoded in two generating functions
\begin{align}
\label{eqn:Fplus} F_+(z) & = \sum_{k=0}^{\infty} m (\frac{n}{2}+k,D)z^k  \\
\label{eqn:Fminus} F_-(z) &= \sum_{k=0}^{\infty} m (-(\frac{n}{2}+k),D)z^k .
\end{align}
It is elementary to show that these power series have radius of convergence at least 1 about $z = 0$.

Now denote the irreducible half spin representations of $Spin(2m)$ by
\begin{align*}
	\rho^+: & Spin(2m) \rightarrow Aut(\Sigma^+_{2m}) \\
	\rho^-: & Spin(2m) \rightarrow Aut(\Sigma^-_{2m}),
\end{align*}
where $\Sigma_{2m}^\pm$ are the positive and negative spinor spaces.
Let $\chi^{\pm}: Spin(2m) \rightarrow \mathbb{C}$ be the character of $\rho^{\pm}$. 
It is shown in \cite{Bar} that the generating functions of the spectral multiplicities have the form
\begin{align}\label{Spmultgen}
F_+ (z) &= \frac{1}{| \Gamma |} \sum_{\gamma \in \Gamma}  \frac{\chi^-(\epsilon(\gamma)) - z \cdot \chi^+ (\epsilon(\gamma))}{det(1_{2m} - z \cdot \gamma)}, \\ \label{Spmultgen2}
F_-(z) &= \frac{1}{| \Gamma |} \sum_{\gamma \in \Gamma} \frac{\chi^+(\epsilon(\gamma)) - z \cdot \chi^- (\epsilon(\gamma))}{det(1_{2m} - z \cdot \gamma)}.
\end{align}

\subsection{The Dirac spectrum of the Poincar\'e sphere}\label{DirPoinSec}

In order to compute explicitly the Dirac spectrum of the Poincar\'e
homology sphere, it suffices then to compute the multiplicities by
computing explicitly the generating functions \eqref{Spmultgen} and \eqref{Spmultgen2}.

Let $\Gamma$ be the binary icosahedral group.  
To carry out our computations, we regard $S^3$ as being the set of unit quaternions, and $\Gamma$ is the following set of 120 unit quaternions:
\begin{itemize}
\item 24 elements are as follows, where the signs in the last group are chosen independently of one another:
\begin{equation}\label{icogrp1}
\{ \pm 1, \pm i , \pm j, \pm k , \frac{1}{2}( \pm 1 \pm i \pm j \pm k) \} .
\end{equation}
\item 96 elements are either of the following form, or obtained by an even permutation of coordinates of the following form:
\begin{equation}\label{icogrp2}
1/2 (0 \pm i \pm \phi^{-1} j \pm \phi k),
\end{equation}
where $\phi$ is the golden ratio.
\end{itemize}
Then $\Gamma$ acts on $S^3$ by left multiplication. Similarly, if $S^3$ is regarded as the unit sphere in $\mathbb{R}^4$, then $SO(4)$ acts on $S^3$ by left multiplication.  In this way, we may identify $a + bi + cj + dk \in \Gamma$, with the following matrix in $SO(4)$:
\[
\left(
\begin{array}{cccc}
a & -b & -c & -d \\
b & a & -d & c \\
c & d & a & -b \\
d & -c & b & a \\
\end{array}
\right)
\]

\subsection{The double cover $Spin(4) \rightarrow SO(4)$}

Let us recall some facts about the double cover $Spin(4) \rightarrow SO(4)$.
Let $S^3_L \simeq SU(2)$ be the group of left isoclinic rotations:
\[
\left(
\begin{array}{cccc}
a & -b & -c & -d \\
b & a & -d & c \\
c & d & a & -b \\
d & -c & b & a \\
\end{array}
\right),
\]
where $a^2 + b^2 + c^2 + d^2 = 1$.
Similarly, let  $S^3_R \simeq SU(2)$ be the group of right isoclinic rotations:
\[
\left(
\begin{array}{cccc}
p & -q & -r & -s \\
q & p & s & -r \\
r & -s & p & q \\
s & r & -q & p 
\end{array}
\right),
\]
where $p^2 +q^2 + r^2 + s^2 = 1$. Then $Spin(4) \simeq S^3_L \times S^3_R$, and the double cover $\Theta: Spin(4) \rightarrow SO(4)$ is given by $(A,B )\mapsto A \cdot B$, where $A \in S^3_L$, and $B \in S^3_R$.  The complex half-spin representation $\rho^-$ is just the projection onto $S^3_L$, where we identify $S^3_L$ with $SU(2)$ via
\[
\left(
\begin{array}{cccc}
a & -b & -c & -d \\
b & a & -d & c \\
c & d & a & -b \\
d & -c & b & a \\
\end{array}
\right)
\mapsto
\left(
\begin{array}{cc}
a -bi & d + ci \\
-d + ci & a + bi 
\end{array}
\right).
\]
The other complex half-spin representation $\rho^+$ is the projection onto $S^3_R$, where we  identify $S^3_R$ with $SU(2)$ via
\[
\left(
\begin{array}{cccc}
p & -q & -r & -s \\
q & p & s & -r \\
r & -s & p & q \\
s & r & -q & p 
\end{array}
\right)^{t}
\mapsto
\left(
\begin{array}{cc}
p - qi & s + ri \\
-s + ri & p + qi 
\end{array}
\right).
\]

\subsection{The spectral multiplicities}

We define our spin structure $\epsilon: \Gamma \rightarrow Spin(4)$ to simply be $A \mapsto (A, I_4)$. It is obvious that this map satisfies $\Theta \circ \epsilon = id_{\Gamma}$.
Therefore, given $\gamma = a + bi + cj + dk \in \Gamma$, we see that 
\begin{align*}
\chi^- (\epsilon(\gamma)) &= 2a \\
\chi^+ (\epsilon(\gamma)) &= 2.
\end{align*}

We then obtain the following result by direct computation of the expressions
\eqref{Spmultgen} and \eqref{Spmultgen2}, substituting the explicit expressions
for all the group elements. This can be done using Mathematica.

\begin{thm}\label{spgenfunPoinc}
Let $S=S^3/\Gamma$ be the Poincar\'e sphere, with the spin structure $\epsilon$ described
here above. The generating functions for the spectral multiplicities of the Dirac operator are
\begin{equation}\label{Fplus}
F_+(z) = -\frac{16(710647 + 317811 \sqrt{5})G^+(z)}{(7 + 3 \sqrt{5})^3 (2207 + 987 \sqrt{5})H^+(z)},
\end{equation}
where 
$$ \begin{array}{rl}
G^+(z) =&  6z^{11} + 18z^{13} + 24z^{15} + 12z^{17} - 2z^{19} \\[2mm]
- & 6z^{21} - 2z^{23} + 2 z^{25} + 4z^{27} + 3z^{29} + z^{31} 
\end{array} $$ and 
$$ \begin{array}{rl}
H^+(z) = & -1 -3 z^{2}  -4z^{4}-2z^{6}+2z^{8}+ 6z^{10} + 9z^{12} + 9z^{14} + 4 z^{16}\\[2mm] 
- &  4 z^{18} - 9 z^{20} -9z^{22}-6z^{24}-2z^{26} + 2z^{28} + 4z^{30} + 3z^{32} + z^{34}, 
\end{array}
$$ and
\begin{equation}\label{Fminus}
F_-(z) = -\frac{1024(5374978561 + 2403763488  \sqrt{5})G^-(z)}{(7 + 3 \sqrt{5})^8 (2207 + 987 \sqrt{5})H^-(z)},
\end{equation}
where 
$$ \begin{array}{rl}
G^-(z) = & 1 + 3z^{2} + 4z^{4} + 2z^{6} - 2z^{8}-6z^{10} \\[2mm]
- & 2z^{12} + 12 z^{14} + 24z^{16} + 18z^{18} + 6z^{20}, 
\end{array}
$$ and 
$$ \begin{array}{rl}
H^-(z) = & -1 -3 z^{2}  -4z^{4}-2z^{6}+2z^{8}+ 6z^{10} + 9z^{12} + 9z^{14} + 4 z^{16} 
\\[2mm] - & 4 z^{18}  - 9 z^{20} - 9z^{22}-6z^{24}-2z^{26} + 2z^{28} + 4z^{30} + 3z^{32} + z^{34}. \end{array} $$
\end{thm}

We can then obtain explicitly the spectral multiplicities from the Taylor coefficients of $F_+(z)$ and $F_-(z)$, as in \ref{eqn:Fplus} and \ref{eqn:Fminus}.

\subsection{The spectral action for the Poincar\'e sphere}

In order to compute the spectral action, we proceed as in the previous cases by
identifying polynomials whose values at the points of the spectrum give the 
values of the spectral multiplicities. We obtain the following result.

\begin{prop}\label{60polys}
There are polynomials $P_k (u)$, for $k = 0, \ldots , 59$, so that $P_k(3/2 + k + 60j) = m(3/2 + k + 60j, D)$ for all $j \in \Z$.  The $P_k(u)$ are given as follows:
\begin{align*}
P_k &= 0, \quad \mathrm{whenever~} k \mathrm{~is~even}. \\
P_1 (u) &= \frac{1}{48} - \frac{1}{20}u + \frac{1}{60}u^2. \\
P_3 (u) &= \frac{3}{80} - \frac{1}{12}u + \frac{1}{60}u^2. \\
P_5 (u) &= \frac{13}{240} - \frac{7}{60}u + \frac{1}{60}u^2. \\
P_7 (u) &= \frac{17}{240} - \frac{3}{20}u + \frac{1}{60}u^2. \\
P_9 (u) &= \frac{7}{80} - \frac{11}{60}u + \frac{1}{60}u^2. \\
\end{align*}
\begin{align*}
P_{11} (u) &= -\frac{19}{48} + \frac{47}{60}u + \frac{1}{60}u^2. \\
P_{13} (u) &= \frac{29}{240} - \frac{1}{4}u + \frac{1}{60}u^2. \\
P_{15} (u) &= \frac{11}{80} - \frac{17}{60}u + \frac{1}{60}u^2. \\
P_{17} (u) &= \frac{37}{240} - \frac{19}{60}u + \frac{1}{60}u^2. \\
P_{19} (u) &= -\frac{79}{240} + \frac{13}{20}u + \frac{1}{60}u^2. \\
\end{align*}
\begin{align*}
P_{21} (u) &= \frac{3}{16} - \frac{23}{60}u + \frac{1}{60}u^2. \\
P_{23} (u) &= -\frac{71}{240} + \frac{7}{12}u + \frac{1}{60}u^2. \\
P_{25} (u) &= \frac{53}{240} - \frac{9}{20}u + \frac{1}{60}u^2. \\
P_{27} (u) &= \frac{19}{80} - \frac{29}{60}u + \frac{1}{60}u^2. \\
P_{29} (u) &= -\frac{59}{240} + \frac{29}{60}u + \frac{1}{60}u^2. \\
\end{align*}
\begin{align*}
P_{31} (u) &= -\frac{11}{48} + \frac{9}{20}u + \frac{1}{60}u^2. \\
P_{33} (u) &= \frac{23}{80} - \frac{7}{12}u + \frac{1}{60}u^2. \\
P_{35} (u) &= -\frac{47}{240} + \frac{23}{60}u + \frac{1}{60}u^2. \\
P_{37} (u) &= \frac{77}{240} - \frac{13}{20}u + \frac{1}{60}u^2. \\
P_{39} (u) &= -\frac{13}{80} + \frac{19}{60}u + \frac{1}{60}u^2. \\
\end{align*}
\begin{align*}
P_{41} (u) &= -\frac{7}{48} + \frac{17}{60}u + \frac{1}{60}u^2. \\
P_{43} (u) &= -\frac{31}{240} + \frac{1}{4}u + \frac{1}{60}u^2. \\
P_{45} (u) &= \frac{31}{80} - \frac{47}{60}u + \frac{1}{60}u^2. \\
P_{47} (u) &= -\frac{23}{240} + \frac{11}{60}u + \frac{1}{60}u^2. \\
P_{49} (u) &= -\frac{19}{240} + \frac{3}{20}u + \frac{1}{60}u^2. \\
\end{align*}
\begin{align*}
P_{51} (u) &= -\frac{1}{16} + \frac{7}{60}u + \frac{1}{60}u^2. \\
P_{53} (u) &= -\frac{11}{240} + \frac{1}{12}u + \frac{1}{60}u^2. \\
P_{55} (u) &= -\frac{7}{240} + \frac{1}{20}u + \frac{1}{60}u^2. \\
P_{57} (u) &= \frac{39}{80} - \frac{59}{60}u + \frac{1}{60}u^2. \\
P_{59} (u) &= -\frac{119}{240} + \frac{59}{60}u + \frac{1}{60}u^2. \\
\end{align*}
\end{prop}

\proof These are computed directly from the Taylor coefficients
of the generating functions of the spectral multiplicities \eqref{Fplus}
and \eqref{Fminus}. \endproof

We then obtain the nonperturbative spectral action for the
Poincar\'e sphere.

\begin{thm}\label{SpActPoinc}
Let $D$ be the Dirac operator on the Poincar\'e homology sphere 
$S=S^3/\Gamma$, with the spin structure $\epsilon: \Gamma \rightarrow Spin(4)$ with 
$A \mapsto (A, I_4)$. Then, for $f$ a Schwartz function, the spectral action is
given by
\begin{equation}\label{SAPo}
\Tr(f(D/\Lambda)) = \frac{1}{60} \left(  \frac{1}{2}\Lambda^3 \widehat f^{(2)}(0) -\frac{1}{8}\Lambda \widehat f(0)  \right),
\end{equation}
which  is precisely 1/120 of the spectral action on the sphere.
\end{thm}

\proof The result follows by applying Poisson summation again, to the
functions $g_j(u)=P_j(u) f(u/\Lambda)$. This
gives, up to an error term which is of the order of $O(\Lambda^{-k})$ for
any $k>0$, the spectral action in the form 
$$ \Tr(f(D/\Lambda)) = \frac{1}{60} \sum_{j=0}^{59} \widehat g_j (0)= 
\frac{1}{60} \int_\R \sum_j P_j(u) f(u/\Lambda) du. $$ 
It suffices then to notice that
$$ \sum_{j=0}^{59} P_j(u) =  \frac{1}{2}u^2-\frac{1}{8}. $$
The result then follows as in the sphere case.
\endproof

\medskip

\subsection{Slow-roll potential in dodecahedral cosmologies}\label{slorolldodecaSec}

The dodecahedral space $S=S^3/\Gamma$, with $\Gamma$ the binary icosahedral
group, also behaves in the same way as the quaternionic space $SU(2)/Q8$ with
respect to the properties of the slow-roll potential and slow-roll parameters. Namely,
the slow-roll potential is a multiple of the potential for the sphere $S^3$ and the
slow-roll parameters are therefore equal to those of the sphere.

\begin{thm}\label{dodecaVphi}
The spectral action for the manifold $S\times S^1$, with $S=S^3/\Gamma$ the
Poincar\'e dodecahedral space, is given by $1/120$ of the spectral action of
$S^3\times S^1$ \eqref{SpActS3S1},
\begin{equation}\label{SpAcDodS1}
\Tr(h(D^2/\Lambda^2)) \sim \frac{\pi}{120} \Lambda^4 a^3 \beta \int_0^\infty u h(u) du -
\frac{\pi}{240} \Lambda^2 a\beta \int_0^\infty h(u) du ,
\end{equation}
up to an error term of the order of $O(\Lambda^{-k})$.
The slow-roll potential $V(\phi)$ obtained by replacing $D^2 \mapsto D^2+\phi^2$ 
is also $1/120$ of the potential for the 3-sphere,
\begin{equation}\label{S3Vphi}
 V(\phi) = \frac{\pi}{120}\Lambda^4 \beta a^3 \cV(\phi^2/\Lambda^2) + \frac{\pi}{240} \Lambda^2 \beta a \cW(\phi^2/\Lambda^2), 
\end{equation} 
with $\cV$ and $\cW$ as in \eqref{VW}. The slow-roll parameters are the
same \eqref{epsilonS3}, \eqref{etaS3} as for the sphere $S^3$.
\end{thm}

\proof We showed in Theorem \ref{SpActPoinc} that the spectral action for the 
Poincar\'e dodecahedral space $S=S^3/\Gamma$ is $1/120$ of the spectral action of the 3-sphere
of radius one. Changing the radius $a$ of the 3-sphere has the effect of changing
$\Lambda \mapsto (\Lambda a)$ in the expression \eqref{SAPo} of
the spectral action, as in the case of the sphere. We then obtain the spectral action
$\Tr(h(D^2/\Lambda^2))$ for the product $S\times S^1$, as in Theorem \ref{SpAcQ8S1thm}
using Poisson summation applied to the functions
$$ g_i(u,v) = 2 P_i(u) h(u^2 (\Lambda a)^{-2} + v^2 (\Lambda \beta)^{-2}), $$
with $P_i(u)$ as in Proposition \ref{60polys}, namely
$$ \Tr(h(D^2/\Lambda^2)) = \sum_{(n,m)\in\Z^2} \sum_{i=0}^{59} g_i (60n +i +\frac{3}{2},
m+\frac{1}{2}) . $$
Then, the Poisson summation formula applied to these functions
shows that the spectral action on
the product of the dodecahedral space by the circle is given by
$$ \Tr(h(D^2/\Lambda^2)) = \frac{1}{60} \sum_i \widehat g_i(0,0) +O(\Lambda^{-k}). $$
We compute this as in the sphere case, using the fact that $\sum_i P_i(u) =u^2/2- 1/8$.
This gives \eqref{SpAcDodS1}, as in the case of the
sphere and of the quaternionic space. The slow-roll potential is then obtained
exactly as in the previous cases.
\endproof
 
\bigskip

\section{Flat cosmologies}

Another very promising candidate for possible non-simply-connected cosmic topologies 
is given by the flat manifolds: flat 3-dimensional tori and their quotients, the Bieberbach
manifolds. 

Simulated CMB skies have been computed for tori and for all the Bieberbach manifolds
in \cite{RWULL}. The method is the same as in the analysis of simulated CMB skies for
spherical space forms of \cite{LWUGL}, \cite{RULW}, namely through the explicit computation
of the spectrum and eigenforms of the Laplacian. In the case of flat tori, the basis given
by planar waves is more directly adapted to the topology, while the basis in spherical
waves is better suited for comparison between simulated and observed CMB sky. So the
analysis of \cite{RWULL} of the Laplace spectra and eigenfunctions uses the transition
between these two bases. The resulting simulated CMB skies are suitable for an
investigation for flat cosmic topologies with the ``circles in the sky" method. A statistical
analysis of distance correlations between cosmic sources, aimed at identifying
possible signatures of cosmic topologies given by flat tori with the method of
``cosmic crystallography" was performed in \cite{GoReTe}.

While an early analysis of the anomalies of the anisotropy spectrum of the CMB (the
quadrupole suppression, the small value of the two-point temperature 
correlation function at large angles, and the quadrupole-octupole alignment)
suggested that flat tori would account for all of these anomalies, if one of the
sides of the fundamental domain is of the order of half the horizon scale,
the more detailed analysis of \cite{OCTZH} excludes this possibility on the basis
of the ``circles in the sky method" and of the $S$-statistic test, measuring 
reflection symmetry.  Nonetheless, the flat tori remain at present one of the most 
promising possible candidates for multiconnected cosmic topologies.

An analysis of how to produce a quadrupole-octupole alignment for a flat 
torus with cubic fundamental domain, depending on the size $\ell$ of the torus,
was given in \cite{ALST}. However, the alignment obtained in this way is not
strong enough to account for the observed anomaly. Comparison with 
candidates such as dodecaredral and octahedral cosmologies
shows that in these spherical topologies one has either no alignment or an
anti-alignment, which appears to favor the flat tori.

We show here that, from the point of view of our model of gravity based on the
spectral action functional, a cosmic topology given by a flat torus generates an
inflation potential and slow-roll parameters that are different from those of
the spherical topologies considered in the previous sections.

\subsection{The spectral action on the flat tori} \label{ToriSpAcSec}

Let $T^3$ be the flat torus $\R^3 / \Z^3$. The spectrum of the Dirac operator, denoted $D_3$,  is given in Theorem 4.1 of \cite{Bar3} as
\begin{equation}\label{torusspec}
\pm 2 \pi \parallel(m,n,p) + (m_0,n_0,p_0)\parallel,
\end{equation}
where $(m,n,p)$ runs through $\Z ^3$.  Each value of $(m,n,p)$ contributes multiplicity 1. The constant vector $(m_0,n_0,p_0)$ depends on the choice of spin structure.

\begin{thm}\label{SpActTori}
The spectral action $\Tr (f(D_3^2/ \Lambda ^2))$ for the torus $T^3=\R^3 / \Z^3$
is independent of the spin structure on $T^3$ and given by
\begin{equation}\label{SpT3D2}
\Tr (f(D_3^2/ \Lambda ^2))=\frac{\Lambda^3}{4 \pi ^3} \int_{\R ^3} f(u^2+ v^2 + w^2) du\,dv\,dw  
+ O(\Lambda^{-k}),
\end{equation}
for arbitrary $k>0$.
\end{thm}

\proof By \eqref{torusspec}, we know the spectrum of $D_3 ^2$ is given by
\[
4 \pi ^2 \parallel (m,n,p) + (m_0,n_0,p_0) \parallel^2,
\]
where $(m,n,p)$ runs through $\Z ^3$, and each value of $(m,n,p)$ contributes multiplicity 2. 

Given a test function in Schwartz space, $f \in \cS(\R )$, the spectral action is then given by
\[
\Tr (f(D_3^2/ \Lambda ^2)) = \sum_{(m,n,p)\in \Z ^3}2 f\left( \frac{4 \pi ^2 ((m+m_0)^2+ (n+n_0) ^2 + (p+p_0)^2)}{\Lambda^2} \right),
\]

In three dimensions, the Poisson summation formula is given by
\[
\sum_{\Z ^3}g(m,n,p) = \sum_{\Z ^3}\widehat{g}(m,n,p),
\]
where the Fourier transform is defined by
\[
\widehat{g}(m,n,p) = \int_{\R ^3} g(u,v,w)e^{-2\pi i (mu + nv + pw)}dudvdw .
\]

If we define 
\begin{equation}\label{guvw}
g(m,n,p) = f\left( \frac{4 \pi ^2 ((m+m_0)^2+ (n+n_0) ^2 + (p+p_0)^2)}{\Lambda^2} \right), 
\end{equation}
and apply the Poisson summation formula, we obtain the following expression for the spectral action:
\begin{align*}
\Tr (f(D_3^2/ \Lambda ^2)) &= 2 \sum_{(m,n,p)\in \Z ^3}\widehat{g}(m,n,p) \\
&= 2\widehat{g}(0,0,0) + O(\Lambda^{-k}) \\
&= 2\int_{\R ^3} f\left( \frac{4 \pi ^2 ((u+m_0)^2+ (v+n_0) ^2 + (w+p_0)^2)}{\Lambda^2} \right) du\, dv\, dw \\ & + O(\Lambda^{-k})\\
&= \frac{\Lambda^3}{4 \pi ^3} \int_{\R ^3} f(u^2+ v^2 + w^2) du\, dv\, dw  + O(\Lambda^{-k})\\.
\end{align*}

The estimate $\sum_{(m,n,p) \neq 0 }\widehat{g}(m,n,p) = O(\Lambda^{-k})$ for arbitrary $k>0$ is elementary, using the fact that $f \in \cS(\R )$.  We observe that the nonperturbative  spectral action is independent of the choice of spin structure.
\endproof

Now let $X = T^3 \times S^1_{\beta}$.  We then compute the spectral action for the 
operator $D_X^2$ as a direct consequence of the previous result.

\begin{thm}\label{SpactT3S1}
On the 4-manifold $X = T^3 \times S^1_{\beta}$, with the flat torus of size $\ell$ and 
with the product Dirac operator
$D_X$ as in \eqref{DXop}, the spectral action is given by
\begin{equation}\label{SpActT3S1}
 \Tr(h(D_X^2/\Lambda^2)) = \frac{\Lambda^4 \beta \ell^3}{4\pi} \int_0^\infty u h(u) du + O(\Lambda^{-k})
\end{equation}
for arbitrary $k>0$.
\end{thm}

\proof  For the operator $D_X^2$, with $D_X$ as in \eqref{DXop} the spectral action
$\Tr(h(D_X^2/\Lambda^2))$ is given by
$$ \sum_{(m,n,p,r)\in \Z^4} 2\,\, h \left( \frac{4\pi^2}{(\Lambda \ell)^2} ( (m+m_0)^2 + (n+n_0)^2
+ (p+p_0)^2) + \frac{1}{(\Lambda \beta)^2} (r+\frac{1}{2})^2 \right). $$
We set
$$ g(u,v,w,y)= 2\,\,  h \left( \frac{4\pi^2}{\Lambda^2} ( u^2+v^2+w^2 ) + \frac{y^2}{(\Lambda \beta)^2} \right). $$
The Poisson summation formula then gives 
$$ \sum_{(m,n,p,r)\in \Z^4} g(m+m_0,n+n_0,p+p_0,r+\frac{1}{2}) =
\sum_{(m,n,p,r)\in \Z^4} (-1)^r \, \widehat g(m,n,p,r). $$
Since we have $h\in \cS(\R)$, we can estimate that the error term
$$ \sum_{(m,n,p,r)\neq (0,0,0,0)} \widehat g(m,n,p,r) $$
is bounded by $O(\Lambda^{-k})$ for arbitrary $k>0$. We then obtain
$$ \Tr(h(D_X^2/\Lambda^2)) = \widehat g(0,0,0,0) + O(\Lambda^{-k}). $$
We have
$$ \widehat g(0,0,0,0) = \int_{\R^4} 2  h \left( \frac{4\pi^2}{(\Lambda \ell)^2} ( u^2+v^2+w^2 ) + \frac{y^2}{(\Lambda \beta)^2} \right) du\,dv\,dw\,dy. $$
This gives 
$$ \frac{\Lambda^4 \beta \ell^3}{4\pi^3} \int_{\R^4} h(u^2+v^2+w^2+y^2)\,  du\,dv\,dw\,dy =
\frac{\Lambda^4 \beta \ell^3 Vol(S^3)}{4\pi^3} \int_0^\infty h(\rho^2) \rho^3 d\rho $$
which gives
$$  \Tr(h(D_X^2/\Lambda^2)) = \frac{\Lambda^4 \beta \ell^3}{2\pi} \int_0^\infty \rho^3 h(\rho^2) d\rho + O(\Lambda^{-k}), $$
from which we obtain \eqref{SpActT3S1}.
\endproof

We now consider the effect of introducing the perturbation $D^2\mapsto D^2 +\phi^2$
in the spectral action. We write as above
$$ \cV(x) = \int_0^\infty u \, (h(u+x)-h(u)) \,du. $$
We then have the following.

\begin{thm}\label{slowrollTori}
The perturbed spectral action on the flat tori is of the form
\begin{equation}\label{toriVphiSpAct}
\Tr(h((D_X^2+\phi^2)/\Lambda^2)) = \Tr(h(D_X^2/\Lambda^2)) + 
\frac{\Lambda^4 \beta \ell^3}{4\pi} \cV(\phi^2/\Lambda^2).
\end{equation}
The corresponding slow-roll potential is of the form
$$ V(\phi)= \frac{\Lambda^4 \beta \ell^3}{4\pi} \cV(\phi^2/\Lambda^2), $$
and the slow-roll parameters are given by
$$ \epsilon =\frac{m_{Pl}^2}{16\pi} \left( \frac{\int_x^\infty h(u) du}{\int_0^\infty u (h(u+x)-h(u)) du }
\right)^2 $$
$$ \eta =  \frac{m_{Pl}^2}{8\pi} \left( \frac{h(x)}{\int_0^\infty u (h(u+x)-h(u)) du} -\frac{1}{2} \left( \frac{\int_x^\infty h(u) du}{\int_0^\infty u (h(u+x)-h(u)) du }
\right)^2 \right). $$
\end{thm}

\proof The result follows directly from \eqref{SpActT3S1} upon writing
$$ \Tr(h((D_X^2+\phi^2)/\Lambda^2)) = \frac{\Lambda^4 \beta \ell^3}{4\pi} \int_0^\infty u h(u) du + O(\Lambda^{-k}) $$
$$ = \frac{\Lambda^4 \beta \ell^3}{4\pi} \int_0^\infty u (h(u+x)-h(u)) du + 
\frac{\Lambda^4 \beta \ell^3}{4\pi} \int_0^\infty u h(u) du + O(\Lambda^{-k}), $$
and computing the slow-roll parameters as in \eqref{epsilonphi} and \eqref{etaphi}.
\endproof

Notice how the absence of the $\cW(\phi^2/\Lambda^2)$ term in the slow-roll
potential for the case of flat tori gives rise to slow-roll parameters that are 
genuinely different from those we computed for the spherical geometries.
This shows that, in noncommutative geometry models of gravity based on
the spectral action functional, there is a nontrivial relation between cosmic
topology (or at least the underlying curvature geometry) and the shape of the 
induced inflation slow-roll potential and parameters.

The case of the other flat geometries, the Bieberbach manifolds, can
be handled with similar techniques, based on the explicit computation
of their Dirac spectra given in \cite{Pfa}.

\bigskip

\section{Geometric engineering of inflation scenarios via Dirac spectra}

If one renounces the assumption of homogeneity and constant curvature,
which reduces the candidate topologies to spherical and flat space forms,
one finds that it is possible to engineer different inflation scenarios, by changing
the slow-roll potential and the resulting slow roll parameters by modifying the
metric on a fixed topology and change accordingly the Dirac spectrum and
the resulting spectral action.

In the spherical examples we computed explicitly in the previous sections the
Dirac spectra tend to have non-trivial multiplicities. These reflect the very
symmetric form of the geometry. On the contrary, it is shown in \cite{Dahl2}
that, for a generic Riemannian metric on a given smooth compact 
3-dimensional manifold $M$, all the Dirac eigenvalues are simple. 

Moreover, the result of \cite{Dahl} shows that, for a given $L>0$ and an
assigned sequence of non-zero real numbers 
\begin{equation}\label{lambdajseqL}
-L < \lambda_1 < \lambda_2 < \lambda_3 < \cdots < \lambda_N < L,
\end{equation}
it is possible to construct, on an arbitrary smooth compact spin 3-manifold $M$,
a Riemannian metric $g$ such that the non-zero spectrum of corresponding Dirac 
operator $D_M$ in the interval $(-L,L)$ consists of the simple eigenvalues
\begin{equation}\label{SpDMgL}
\Spec(D_M) \cap ((-L,L)\smallsetminus \{ 0 \}) = \{ \lambda_j \}_{j=1,\ldots,N} .
\end{equation}
The way to obtain a Dirac spectrum with these properties is to start with a metric on $M$
for which the dimension of the kernel of the Dirac operator is minimal, compatibly with 
the constraint given by the index theorem. By rescaling this metric one ensures that
no other eigenvalue occurs in an interval $(-3L,3L)$. One then performs a connected
sum with $N$ copies of $S^3$, so that the resulting manifold is still topologically the
same as $M$. One endows each of the 3-spheres with a Berger metric as in   
\cite{Hitchin}, scaled so that the interval $(-2L,2L)$ contains only one eigenvalue of
the Dirac operator. Then applying a surgery formula one obtains the desired eigenvalues
for the Dirac spectrum on the connected sum manifold, \cite{Dahl}. 

For simplicity, to avoid handling separately a possible kernel, let us consider here a variant of the
spectral action where one only sums over the non-zero spectrum of $D$. We write this
as 
\begin{equation}\label{primeSpAct}
\Tr'(f(D/\Lambda)):= \sum_{\lambda\in \Spec(D)\smallsetminus \{0\}} f(\lambda/\Lambda).
\end{equation}

We then have the following result, which allows us to construct, on a given 3-manifold
a metric with prescribed spectral action.

\begin{lem}\label{arprogSpAct}
Let $M$ be a compact smooth 3-manifold, with a given spin structure. Let $f$ be a
smooth function, compactly supported inside an interval $[-L/\Lambda,L/\Lambda]$. 
Then, for any given $\lambda >0$, there is a metric $g_{\lambda,L}$ on $M$ such 
that the spectral action for the resulting Dirac operator is
\begin{equation}\label{primeSpActgL}
\Tr'(f(D/\Lambda))= \frac{\Lambda}{\lambda} \hat f(0) + O(\Lambda^{-k}),
\end{equation}
for arbitrary $k>0$.
\end{lem}

\proof Let $\lambda_n =\eta + n \lambda$ be a progression indexed
by the integers $n\in \Z$, with $\lambda>0$ and $\eta \neq 0$. Let
$\{ \lambda_{n_0+j} \}_{j=1,\ldots, N}$ be the points of this sequence
that lie in the interval $(-L,L)$. We assume that $\lambda_n \neq \pm L$
for all $n$. Using the method of \cite{Dahl} we construct, by taking
connected sums with Berger spheres, a metric on $M$ for which the
Dirac operator $D$ has $\Spec(D)\cap ((-L,L)\smallsetminus \{0\})$
given by the simple eigenvalues $\lambda_{n_0+j}$, with $j=1,\ldots, N$.
For a test function supported in $[-L/\Lambda,L/\Lambda]$ we then have
$$ \Tr'(f(D/\Lambda))= \sum_{n\in \Z} f(\lambda_n/\Lambda). $$
We can then use the Poisson summation formula 
$$ \sum_{n\in \Z} g(\eta + n \lambda) =\sum_{n\in \Z} \frac{1}{\lambda} \exp(\frac{2\pi i n \eta}{\lambda}) \widehat g(\frac{n}{\lambda}) $$
to $g(u)=f(u/\Lambda)$. We estimate as in the previous cases that
$$ \left| \sum_{n\neq 0} \frac{1}{\lambda} \exp(\frac{2\pi i n \eta}{\lambda}) \widehat g(\frac{n}{\lambda}) \right| \leq O(\Lambda^{-k}), $$
for arbitrary $k>0$, so that we are left with the term
$$ \frac{1}{\lambda} \widehat g(0) = \frac{\Lambda}{\lambda} \widehat f(0). $$
\endproof

Notice, for example, that we can apply this result starting from any one of the spherical
topologies we analyzed in the previous sections. In such cases one starts
with the round metric, so one does not have a kernel to worry about. One
then scales it so as not to have any other eigenvalue in the interval $(-3L,3L)$
and proceeds to modify the metric by taking connected sums with the
Berger spheres to insert the desired eigenvalues in the interval $(-L,L)$.
Thus, on a given underlying topology one can significantly alter the form
of the spectral action by this method, at the cost of no longer having
a homogeneous metric. We now show the effect this operation has on the inflation
slow-roll potential, even though such non-homogeneous metrics are clearly
less interesting in terms of candidate cosmologies. 

We now see how this procedure can be used to construct different possible
slow-roll potentials. 

Let $(\lambda_n, P)$ denote the following data:
\begin{itemize}
\item A progression $\lambda_n =\eta + n \lambda$, for $n\in \Z$, with $\lambda>0$ and $\eta \neq 0$. 
\item  A polynomial $P(u)=\alpha u^2 + \gamma$ with the property that $P(\lambda_n)=m_n$
is a non-negative integer, for all $n$. 
\end{itemize}

In the following, as above, we assume that either we start from a manifold $M$ with a metric
for which $D_M$ has trivial kernel, or else we modify the spectral action on $M\times S^1$
to count only the non-zero part of the spectrum of $D_M$.

\begin{prop}\label{buildSpAct}
Let $M$ be a compact smooth 3-dimensional manifold endowed with a spin structure. 
Given a smooth compactly supported test function $h$ such that $h\equiv 1$ on an
interval $[-T,T]$ and decays rapidly to zero outside of this interval, and given  
a choice of data $(\lambda_n, P)$ as above, there exists a Riemannian metric $g$ 
on $M$ such that the resulting Dirac operator $D$ of the form  \eqref{DXop} on
$M \times S^1_\beta$ has spectral action
\begin{equation}\label{SpActMS1build}
\Tr (h(D^2/\Lambda^2)) =\frac{ \pi \Lambda^4 \beta \alpha }{\lambda} \int_0^\infty 
u\, h(u)\, du + \frac{ 2 \pi \Lambda^2 \beta  \gamma}{\lambda} \int_0^\infty h(u) \, du
+ O(\Lambda^{-k}),
\end{equation}
for arbitrary $k>0$.
\end{prop}

\proof Consider the sequence of non-negative integers $m_n =P(\lambda_n)$.
Let $\Omega \subset \Z^2$ be the set of pairs $(n,m)$ such that
$$ x_{n,m}(\Lambda,\beta) := \frac{\lambda_n^2}{\Lambda^2} + \frac{(m+1/2)^2}{(\Lambda
\beta)^2} \in (0,T). $$
We can assume that all other points of the form $x_{n,m}$, for $(n,m)\notin \Omega$ 
lie outside of the support of $h$. Let $(-L,L)$ be an interval that contains all the
points $\lambda_n$ for which the set of $m\in \Z$ with $(n,m)\in \Omega$ is non-empty. 
For all $n\in \Z$ such that the set of $(n,m)\in \Omega$ is nonempty, choose
sufficiently small, non-intersecting open intervals 
$U_{n,\epsilon}=(\lambda_n-\epsilon,\lambda_n+\epsilon)$
around the value $\lambda_n$, such that all $\lambda \in U_{n,\epsilon}$ have the
property that $\frac{\lambda^2}{\Lambda^2} + \frac{(m+1/2)^2}{(\Lambda
\beta)^2} \in (0,T)$, for all $m$ such that $(n,m)\in \Omega$. Then choose
$m_n$ points $\lambda_{n,1} < \lambda_{n,2} < \cdots < \lambda_{n,m_n}$
inside $U_{n,\epsilon}$. By the construction of \cite{Dahl}, by taking connected
sums with suitable Berger spheres, we can obtain on $M$ a metric for which the
Dirac operator $D_M$ has spectrum satisfying 
$\Spec(D_M)\cap ((-L,L)\smallsetminus \{0\})=\{ \lambda_{n,j} \}$ with $n=n_0,\ldots,n_0+N$
and $j=1,\ldots, m_n$. Let $D$ be the corresponding Dirac operator \eqref{DXop} on
$M \times S^1_\beta$. The spectral action $\Tr (h(D^2/\Lambda^2))$ is computed by
$$ \sum_{n,j} 2\, h( \lambda_{n,j}^2 \Lambda^{-2} + (m+1/2)^2 (\Lambda \beta)^{-2}) . $$
Given the construction of the $\lambda_{n,j}$ above, this is also equal to 
$$ \sum_{(n,m)\in \Z^2} 2 P(\lambda_n)  h( \lambda_n^2 \Lambda^{-2} + (m+1/2)^2 (\Lambda \beta)^{-2}). $$ 
We let $g(u,v)= 2P(u) h( u^2 \Lambda^{-2} + v^2 (\Lambda \beta)^{-2})$ and we obtain,
by the Poisson summation formula,
$$ \sum_{(n,m)\in \Z^2} g(n \lambda + \eta, m + \frac{1}{2}) =
\sum_{(n,m)\in \Z^2} (-1)^m \exp(\frac{2\pi i n \eta}{\lambda}) \, \frac{1}{\lambda} 
\widehat g(\frac{n}{\lambda},m). $$
Estimating as before the sum of terms with $(n,m)\neq (0,0)$ to be bounded by
$O(\Lambda^{-k})$ for arbitrary $k>0$, this gives
$$ \Tr (h(D^2/\Lambda^2)) = \frac{1}{\lambda} \widehat g(0,0) + O(\Lambda^{-k}), $$
where we then have
$$ \widehat g(0,0) = \int_{\R^2} 2 (\alpha u^2 +\gamma)\, 
h( u^2 \Lambda^{-2} + v^2 (\Lambda \beta)^{-2}) \, du \, dv $$
$$ = \Lambda^4 \beta \alpha \int_{\R^2} (u^2+v^2)\, h(u^2+v^2)\, du\, dv +
2 \Lambda^2 \beta \gamma \int_{\R^2} \, h(u^2+v^2)\, du\, dv $$
$$ = \Lambda^4 \beta \alpha 2\pi \int_0^\infty \rho^3 h(\rho^2)\, d\rho 
+ \Lambda^2 \beta \gamma 4\pi \int_0^\infty h(\rho^2)\, d\rho $$
$$ = \Lambda^4 \beta \alpha \pi \int_0^\infty u \, h(u)\, du +
\Lambda^2 \beta \gamma 2\pi \int_0^\infty h(u)\, du. $$
\endproof

We then have the following result, which shows that altering the spatial 
metric on suitable bubbles (the Berger spheres with which
one performs a connected sum) consequently alters the form of the
inflation potential and slow-roll parameters induced by the spectral action.

\begin{cor}\label{slowrollbuild}
For a 3-manifold $M$ with a metric constructed as in Proposition \ref{buildSpAct}
above, the induced slow-roll potential has the form
$$ V(\phi)  = \frac{\pi \Lambda^4 \alpha \beta}{\lambda} \cV(\phi^2/\Lambda^2) 
- \frac{2 \pi \Lambda^2 \gamma \beta}{\lambda} \cW(\phi^2/\Lambda^2), $$
with $\cV$ and $\cW$ as in \eqref{VW}. The corresponding slow-roll parameters
are given by
$$ \epsilon(x) = \frac{m_{Pl}^2}{16\pi} \left( \frac{\alpha \Lambda^2 \cV^\prime(x) -
2 \gamma \cW^\prime(x)}{ \alpha \Lambda^2 \cV(x) - 2 \gamma \cW(x)} \right)^2 $$
$$ \eta(x) = \frac{m_{Pl}^2}{8\pi} \left( \frac{\alpha \Lambda^2 \cV^{\prime\prime}(x)
- 2 \gamma \cW^{\prime\prime}(x)}{\alpha \Lambda^2 \cV(x) - 2 \gamma \cW(x)} 
- \frac{1}{2} \left( \frac{\alpha \Lambda^2 \cV^\prime(x) -
2 \gamma \cW^\prime(x)}{ \alpha \Lambda^2 \cV(x) - 2 \gamma \cW(x)} \right)^2 \right). $$
\end{cor}

\proof This follows directly from the previous result, by computing
$\Tr(h(D^2+\phi^2)/\Lambda^2)$ in the same way as in the previous cases.
\endproof

Notice that, after rotating back to Lorentzian signature with a metric of the Friedmann
form \eqref{Friedmannds}, the factor $\Lambda$ in the slow-roll parameter appears 
in fact multiplied by the scale factor $a(t)$ of the Friedmann metric, which gives a 
constant term by the relation $\Lambda(t)\sim 1/a(t)$, as in the previous cases.
One is left with the freedom of modifying the slow-roll parameters by changing the
modified metric on the Berger spheres and correspondingly affecting the values of
the parameters $\alpha$ and $\gamma$.

One can also obtain potentials of a more general form, if one constructs, via
the same method, spectra that are only partially given by arithmetic progressions.
An example of this sort is computed explicitly in the appendix: it gives rise to
a genuinely different shape of the potential $V(\phi)$.

\section{Conclusions}

In models of high-energy physics based on noncommutative geometry, the
spectral action functional of \cite{ChCo} is proposed as an action functional
for gravity, or for gravity coupled to matter when additional noncomumutative
extra dimensions are introduced in the geometry of the model. We concentate
here on the purely gravitational part of the model, without noncomumutative 
extra dimensions, and we compute the explicit {\em nonperturbative} form 
of the spectral action functional for three among the more likely 
candidates for the problem of cosmic topology: the quaternionic space
$SU(2)/Q8$ and the Poincar\'e
dodecahedral space $S^3/\Gamma$, with $\Gamma$ the binary icosahedral
group, and the flat tori. We show that when one computes the
spectral action for the 4-dimensional manifold obtained by Wick rotating
and compactifying the corresponding space-time to a product of the
given 3-manifold by a circle, one obtains as non-perturbative effect
a slow-roll potential for a field $\phi$ coming from perturbations
$D^2 \mapsto D^2 +\phi$ of the Dirac operator of the 4-dimensional
geometry. We compute the slow-roll parameters for the resulting
slow-roll potential $V(\phi)$ and show that they make sense when
rotating back to the original Minkowskian spacetime. We see that,
in the case of the quaternionic
and the dodecahedral space, the slow-roll parameters are the same as for the ordinary
case of the sphere $S^3$, while in the case of the flat tori the potential one
obtains in this way behaves significantly 
differently from the spherical cases. This shows that cosmological models based on
noncommutative geometry predict that different candidate cosmic topologies
may give rise to different inflation scenarios,
and different values for testable cosmological parameters.

\bigskip
\bigskip

\section{Appendix: Lens spaces, a false positive}\label{LensFalseSec}

Lens spaces are quotients of the sphere $S^3$ by the action of a finite cyclic group $\Z/N\Z$.
They have been considered among the candidate cosmic topologies, especially in \cite{URLW},
which shows simulated CMB maps for lens spaces and computes the expected CMB anisotropies
for some of these topologies. The surprising result of the analysis of \cite{URLW} is that instead
of finding an increasingly suppressed quadrupole with increasing $N$, the low multipoles are
enhanced instead of being suppressed for large $N$ Thus, the simulated power spectra of \cite{URLW}
suggest that to maintain consistency with the WMAP data, one cannot exceed the range $N\leq 15$.
On the other hand, in the same work \cite{URLW} the lens space case is analyzed from the point
of view of the ``circles in the sky" method and it is shown that potentially detectable periodicities
(matching circles) would appear only in the range $N> 7$.

\subsection{The trouble with the Dirac spectrum on lens spaces}

Consider in particular lens spaces 
$\cL_N =SU(2)/\Z_N$, with $N\geq 3$, which are quotients of the 
sphere $SU(2)=S^3$ by the action of the finite cyclic group $\Z_N=\Z/N\Z$ acting on
$SU(2)\subset \C^2$ by
\begin{equation}\label{lensact}
 \left( \begin{array}{cc} \omega & 0 \\ 0 & \omega^{-1} \end{array} \right), \ \ \  \text{ with } \ \ 
 \omega^N=1 . 
\end{equation} 
For these lens spaces, B\"ar gave an explicit
computation of the Dirac spectrum in \cite{Bar2}. The result states that, for the 
canonical spin structure, the spectrum is of the form
\begin{equation}\label{lensspectrum}
\begin{array}{rl} 
(i) &\quad -iN - \frac{1}{2} ,~ \mathrm{with~multiplicity} ~2iN,~i=0,1,2,\ldots\\[3mm]
(ii) &\quad -\frac{1}{2} \pm m, ~ \mathrm{with~multiplicity}~m,\\[3mm]
& m = 2,3,\ldots,-(m+1)< iN \leq (m-2) 
\end{array}
\end{equation} 
where we have taken on the sphere $S^3$ the round metric of radius one. 
In the even case $N=2N'$, there is also a second spin structure for which the
Dirac spectrum is given in \cite{Bar2} as
\begin{equation}\label{SpLN2spin}
\begin{array}{ll}
(i) & -(N' + iN) - \frac{1}{2} \  \text{ with multiplicity } \,\, 2(N' + iN),~i=0,1,2,\ldots \\[2mm]
(ii) & -\frac{1}{2} \pm m \ \text{ with multiplicity } \,\,  m,\\
&  m = 2,3,\ldots \quad 1-m < iN + \frac{N}{2} \leq m-2. 
\end{array} 
\end{equation}

Unfortunately, this result of \cite{Bar2} appears to be incorrect, as we discuss
in \S \ref{LensTrueSec} below.

However, we still show here what the spectral action and slow-roll potential
would be for a manifold with Dirac spectrum as above, because the computation
itself exhibits some interesting features that we have not encountered in the other
spherical and flat examples and that may be useful in different contexts, for manifolds
whose Dirac spectrum only partially decomposes as a union of arithmetic
progressions. 

We show that the incorrect calculation of the Dirac spectrum of $\cL_N$
of \cite{Bar2} leads to a ``false positive" result of a spherical cosmic topology which
gives rise to an inflation scenario different from the simply connected case.
However, as we show in \S \ref{LensTrueSec} below, with the correct
calculation of the Dirac spectrum for the lens spaces, the inflation potential
is in fact again the same as for the case of the sphere, just as in the other
spherical cases we computed in this paper. 

\medskip

\subsection{Multiplicities, first case}

We consider here the problem of computing the explicit nonperturbative
form of the spectral action for an operator $D$ with spectrum of the form
\eqref{lensspectrum}.

We start by writing the multiplicities in a more convenient form.
The multiplicity in row $(i)$ is already in a nice form. To handle row $(ii)$, 
we need to break it up into the subsets corresponding to each equivalence class 
$m \equiv j$ (mod $N$), where $j \in \{0,1,\ldots, N-1\}$.

In order to determine the multiplicity of $-1/2 \pm m$, it is convenient to replace the upper and lower bounds of $-(m+1)< iN \leq (m-2)$ with the smallest and largest values of $iN$ which satisfy the inequality.

\begin{lem}\label{multipLN}
The multiplicity of $-1/2 \pm m$ is given by
\begin{equation}\label{multLN}
\begin{array}{ll}
\displaystyle{\frac{2m(m-j)}{N}}
& \text{ for }  m \equiv j \mod N,  \text{ with } j = 0,1 \\[3mm]
\displaystyle{\frac{2m^2 - 2mj + mN}{N} }
& \text{ for } m \equiv j \mod N,  \text{ with }  j = 2,3,\ldots, N-1.
\end{array}
\end{equation}
\end{lem}

\proof
We first look at the case where $m \equiv j$ (mod $N$), $j = 0,1$.
In this case, the bound $-m -1 < iN \leq m-2$ can be replaced by
$$ j-m \leq iN \leq m - j - N, $$
by adding $j + 1$ to the left hand side and subtracting $N + j - 2$ from the right hand side.  
What matters is that $0< j+1 \leq N$, and $0 \leq N + j -2 < N$.

If $m = kN + j$, where $k= 0,1,2,\ldots$, then we see that there are $2k$ values of $i$ which satisfy the inequality, and hence $-1/2 \pm m$ has multiplicity $2km = (2m(m-j))/N$. Notice that when $m$ is $0$ or $1$, this formula gives us a multiplicity of zero, which is good, since in row $(ii)$, the index $m$ begins at $m=2$.

We then consider the case $m \equiv j$ (mod $N$), $j = 2,3,\ldots, N-1$. One can
replace the upper and lower bounds $-m -1 < iN \leq m-2$ by
$$ j - m \leq iN \leq m - j, $$
by adding $j + 1$ to the left hand side, and subtracting $j-2$ from the right hand side.
One has $0 < j+1 \leq N$, and $0 \leq j-2 < N$.

If $m = kN + j$, $k = 0,1,2,\ldots$, then we see that there are $2k+1$ values of $i$ which satisfy the inequality, and hence  $-1/2 \pm m$ has multiplicity $(2k + 1) m = (2m^2 - 2mj + mN)/N$.
\endproof

\subsection{The spectral action and the Poisson formula}

One sees then that, unlike the cases of the spherical topologies
we analyzed before, one cannot simply write the whole spectrum \eqref{lensspectrum}
as a union of arithmetic progressions indexed over the integers. However,
it is still possible to extend the positive and the negative part of the spectrum,
separately, to unions of such arithmetic progressions. 

This provides us with a different method, still based on the Poisson summation
formula, to compute the spectral action, which may turn out to be useful
in other cases. This is our main reason for including the full calculation
here, despite the fact that it does not give the correct answer for lens
spaces. 

One finds that the multiplicities, for the positive and the negative parts of the 
spectrum, can be interpolated by polynomials, in the following way.

\begin{lem}\label{posmultLN}
For $m>0$, the multiplicity of $-1/2 +m$, when $m\equiv i$ mod $N$, is given by $P_i^+(-1/2+m)$,
with $P_i^+$ the polynomials
\begin{equation}\label{PiplusLN}
\begin{array}{rl}
P_0^+ (u) =& \frac{2}{N}u^2 + \frac{2}{N}u+\frac{1}{2N} \\[2mm]
P_1^+ (u) =& \frac{2}{N}u^2- \frac{1}{2N}\\[2mm]
P_j^+ (u) =& \frac{2}{N}u^2 +\frac{2-2j + N}{N}u+\frac{1-2j + N}{2N}, \quad j = 2,3,\ldots, N-1,
\end{array}
\end{equation}
For $m\geq 0$, the multiplicity of $-1/2 -m$, when $m\equiv \ell$ mod $N$ is given by
$P^-_\ell(-1/2-m)$, with $P_\ell^-$ the polynomial 
\begin{equation}\label{PiminusLN}
\begin{array}{rl}
P_0^- (u) =& \frac{2}{N}u^2+ \frac{2}{N}u+\frac{1}{2N} \\[2mm]
P_1^- (u) =& \frac{2}{N}u^2+ \frac{4}{N}u+\frac{3}{2N}\\[2mm]
P_j^- (u) =& \frac{2}{N}u^2 + \frac{2+ 2j - N}{N}u+\frac{1 + 2j - N}{2N}, \quad j = 2,3,\ldots, N-1.
\end{array}
\end{equation}
The multiplicity of $-iN-1/2$, for $i\geq 0$, is given by $P^-(-iN-1/2)$, with
\begin{equation}\label{PminusLN}
P^-(u) =-2u -1.
\end{equation}
\end{lem}

\proof This follows directly from the expressions for the multiplicities given in Lemma
\ref{multipLN} above. \endproof

One can then make the following observation on computing the spectral action.

\begin{lem}\label{SpactpmLN}
Let $D$ be an operator with spectrum  \eqref{lensspectrum}.
Given a Schwartz function $f$, there are Schwartz functions  $f_+$ and $f_-$, 
respectively supported on the positive and negative reals, with the property
that $f=f_+ + f_-$ on $(-\infty,-\alpha]\cup [\alpha,\infty)$, with $I_\alpha=(-\alpha,\alpha)$ 
an interval with $\Sp(D)\cap I_\alpha =\emptyset$. The spectral action for $D$ is 
then computed by
\begin{equation}\label{fpmScAct}
\Tr(f(D/\Lambda)) = \Tr(f_+(D/\Lambda))+\Tr(f_-(D/\Lambda)).
\end{equation}
\end{lem}

\proof One observes from \eqref{lensspectrum} that there is a gap in the spectrum of
$D$ around zero. Thus, it is possible to replace the function $f$ with a pair of Schwartz
functions $f_+$ and $f_-$, which are, respectively, equal to $f$ on the positive and 
negative parts of the spectrum and that have support contained only in the positive
or negative reals. Since the values of $f$ and $f_+ + f_-$ on an open neighborhood 
of the spectrum are the same, the value of the spectral action is unchanged.
\endproof

We can now compute the two terms on the right hand side of \eqref{fpmScAct}.

\begin{thm}\label{posSpAcLN}
Let $f_+$ be a Schwartz function supported on the positive reals, chosen as
in Lemma \ref{SpactpmLN}. Then, for an operator $D$ with spectrum of the 
form \eqref{lensspectrum} one has
\begin{equation}\label{fplusSpAc}
\Tr(f_+(D/\Lambda)) = \frac{1}{N} \left( 2 \Lambda^3 \widehat f_+^{(2)}(0)  + \Lambda^2 \widehat f_+^{(1)}(0) \right) + \epsilon_+(\Lambda),
\end{equation}
where the error term is of order $\epsilon_+(\Lambda)=O(\Lambda^{-k})$, for any $k>0$,
and where $\widehat f_+^{(k)}$ is the Fourier transform of $v^k f_+(v)$, as above.
\end{thm}

\proof We define $g_j ^+ (u) = P_j^+(u) f_+(u/ \Lambda)$, with $P_j^+$ as in \eqref{PiplusLN},
for $j=0,\ldots,N-1$, so that we can write the spectral action with test function $f_+$ in the
form 
\begin{equation}\label{SpActLNplusgj}
\Tr(f_+(D/\Lambda)) = \sum_{k\in \Z} \sum_{j=0}^{N-1} g_j^+(-1/2+kN+j).
\end{equation}
In fact, extending the sum to $m\in \Z$ does not change anything, since 
all terms with $m\leq 0$ fall outside of the support of $f_+$. We can then
apply the Poisson summation formula to compute this expression. The analog
of \eqref{g4nPoisson} now gives
\begin{equation}\label{gjplusPoisson}
\sum_{k\in \Z}  g_j^+(kN+(2j-1)/2) = \sum_{k\in \Z} \frac{1}{N}\,\, \exp\left(\frac{(2j-1)\pi i k}{N}\right) 
 \,\, \widehat g_j^+(\frac{k}{N}).
\end{equation}

The same argument used in \cite{uncanny} to estimate the remainder term
applies here to give, for any $k>0$,
$$ \sum_{k\neq 0}  \frac{1}{N} | \widehat g_j^+(\frac{k}{N}) | \leq O(\Lambda^{-k}), $$
so that \eqref{SpActLNplusgj} can then be written as
\begin{equation}\label{SpActLNplushatgj}
\Tr(f_+(D/\Lambda)) = \frac{1}{N} \sum_{j=0}^{N-1} \widehat g_j^+(0) + O(\Lambda^{-k}).
\end{equation}

We then obtain the following values for $\widehat g_j^+(0)$, using the
form \eqref{PiplusLN} of the polynomials $P^+_j$:
\begin{equation}\label{hatgj0}
\begin{array}{rl}
\widehat{g}_0^+ (0) =&\displaystyle{
 \frac{2}{N} \Lambda^3 \widehat f_+^{(2)}(0) + \frac{2}{N} \Lambda^2 \widehat f_+^{(1)}(0) +  \frac{1}{2N} \Lambda \widehat f_+(0) }  \\[2mm]
\widehat{g}_1^+ (0)  =& \displaystyle{ \frac{2}{N} \Lambda^3 \widehat f_+^{(2)}(0) -  
\frac{1}{2N} \Lambda \widehat f_+(0) }
\\[2mm]
\widehat{g}_j^+ (0) =& \displaystyle{ \frac{2}{N}\Lambda^3 \widehat f_+^{(2)}(0) +  \frac{2-2j + N}{N} \Lambda^2 \widehat f_+^{(1)}(0) +  \frac{1-2j + N}{2N} \Lambda \widehat f_+(0)} , \\ & j = 2,3,\ldots, N-1
\end{array}
\end{equation}
This then gives
$$
\frac{1}{N} \sum_{j=0}^{N-1} \widehat{g_j^+}(0)  = \frac{1}{N} \left( 2 \Lambda^3 \widehat f_+^{(2)}(0) + 
\Lambda^2 \widehat f_+^{(1)}(0) \right),
$$
while the terms with $\widehat f_+(0)$ in this case add up to zero, since 
$$ \frac{2}{N} + \sum_{j=2}^{N-1} \frac{2 -2j +N}{N} 
= 1, \ \ \ \text{ and } \ \ \
 \sum_{j=2}^{N-1} \frac{1-2j + N}{2N} 
=0. $$
\endproof

The argument for the term with $f_-$ is similar. We have the following result.

\begin{thm}\label{negSpAcLN}
Let $f_-$ be a Schwartz function supported on the negative reals, chosen as
in Lemma \ref{SpactpmLN}. Then, for $D$ an operator with spectrum \eqref{lensspectrum} 
one has
\begin{equation}\label{fminusSpAc}
\Tr(f_-(D/\Lambda)) = \frac{1}{N} \left( 2 \Lambda^3 \widehat f_-^{(2)}(0)  + \Lambda^2 \widehat f_-^{(1)}(0) \right) + \epsilon_-(\Lambda)
\end{equation}
with the error term $\epsilon_-(\Lambda)=O(\Lambda^{-k})$, for any $k>0$.
\end{thm}

\proof We set $g_j^- ( u) = P_j^- (u) f_-(u/\Lambda)$, with $P_j^-$ as in \eqref{PiminusLN},
and $g^-(u)=P^-(u) f_-(u/\Lambda)$, with $P^-$ as in \eqref{PminusLN}. Then we can write
the spectral action on $\cL_N$, with the Schwartz function $f_-$, in the form
\begin{equation}\label{SpActLNminusgj}
\Tr(f_-(D/\Lambda)) = \sum_{k\in \Z} \left( g^-(kN-1/2) + \sum_{j=0}^{N-1} g_j^-(-1/2+kN-j) \right).
\end{equation}
Then one can again use the Poisson summation formula 
\begin{equation}\label{gjminusPoisson}
\sum_{k\in \Z}  g_j^-(kN-(2j+1)/2) = \sum_{k\in \Z} \frac{1}{N}\,\, \exp\left(\frac{-(2j+1)\pi i k}{N}\right) 
 \,\, \widehat g_j^-(\frac{k}{N})
\end{equation}
and
\begin{equation}\label{gminusPoisson}
\sum_{k\in \Z} g^-(kN-1/2) = \sum_{k\in \Z} \frac{1}{N}\,\, \exp\left(\frac{-\pi i k}{N}\right) 
\widehat g^-(\frac{k}{N}),
\end{equation}
and an estimate of the error terms 
$$ \sum_{k\neq 0}  \frac{1}{N} | \widehat g_j^-(\frac{k}{N}) | \leq O(\Lambda^{-k}) \ \ 
\text{ and } \ \ 
\sum_{k\neq 0}  \frac{1}{N} | \widehat g^-(\frac{k}{N}) | \leq O(\Lambda^{-k}), $$
as in \cite{uncanny} to write \eqref{SpActLNminusgj} as 
\begin{equation}\label{SpActLNminushatgj}
\Tr(f_-(D/\Lambda)) = \frac{1}{N}\left( \widehat g^-(0)+ \sum_{j=0}^{N-1} \widehat g_j^-(0) \right) + O(\Lambda^{-k}).
\end{equation}
One can then compute these values using the explicit form of the polynomials $P_j^-$ and $P^-$
of  \eqref{PiminusLN} and \eqref{PminusLN} and one obtains
\begin{equation}\label{hatgjgminus0}
\begin{array}{rl}
\widehat g_0^- (0) =& \displaystyle{\frac{2}{N}\Lambda^3 \widehat f_-^{(2)}(0)  + \frac{2}{N}\Lambda^2 \widehat f_-^{(1)}(0) +  \frac{1}{2N} \Lambda \widehat f_-(0)} \\[3mm]
\widehat g_1^- (0) =& \displaystyle{ \frac{2}{N}\Lambda^3 \widehat f_-^{(2)}(0) + \frac{4}{N} \Lambda^2 \widehat f_-^{(1)}(0) +  \frac{3}{2N} \Lambda \widehat f_-(0)} \\[3mm]
\widehat g_j^- (0) =&\displaystyle{ \frac{2}{N}\Lambda^3 \widehat f_-^{(2)}(0) + \frac{2 + 2j - N}{N} \Lambda^2 \widehat f_-^{(1)}(0) +  \frac{1 + 2j - N}{2N} \Lambda \widehat f_-(0),} \\
&  j = 2,3,\ldots, N-1\\[3mm]
\widehat g^- (0) =& - 2\Lambda^2 \widehat f_-^{(1)}(0) - \Lambda \widehat f_-(0) .
\end{array}
\end{equation}
Thus, since
$$ \frac{2}{N} + \frac{4}{N} + \sum_{j=2}^{N-1} \frac{2+2j-N}{N} -2 
= 1 \ \ \text{ and } \ \ 
\frac{1}{2N} + \frac{3}{2N} +  \sum_{j=2}^{N-1} \frac{1+2j-N}{2N} -1 
=0, $$
one then has
\begin{equation}\label{sumhatgjg0}
\widehat g^-(0)+ \sum_{j=0}^{N-1} \widehat g_j^-(0) =
 2 \Lambda^3 \widehat f_-^{(2)}(0) + \Lambda^2 \widehat f_-^{(1)}(0) .
\end{equation}
Thus, one obtains \eqref{fminusSpAc}.
\endproof

This gives a complete nonperturbative calculation of the spectral action as follows.

\begin{thm}\label{SpActLN}
The spectral action for an operator $D$ with spectrum \eqref{lensspectrum} 
is given by
\begin{equation}\label{fSpAcLN}
\Tr(f(D/\Lambda)) \sim \frac{1}{N} \left( 2 \Lambda^3 (\widehat f_+^{(2)}(0)+\widehat f_-^{(2)}(0))  +  \Lambda^2 (\widehat f_+^{(1)}(0)+\widehat f_-^{(1)}(0)) \right) ,
\end{equation}
up to an error term of the order of $O(\Lambda^{-k})$.
\end{thm}

\proof This follows directly from Lemma \ref{SpactpmLN} and Theorems \ref{posSpAcLN} and
\ref{negSpAcLN}. \endproof

In particular, one is especially interested in the case where the function $f$ is a Schwartz
function that approximates a cutoff function on an interval $[-\alpha,\alpha]$. In this case,
$f$ is an even function and one can assume that the two functions $f_+$ and $f_-$ can
be chosen to be mirror images, so that $f_+(x)=f_-(-x)$. 

\begin{cor}\label{SpActLNeven}
Let $f$ be an even Schwartz function such that the $f_+$ and $f_-$ of Lemma \ref{SpactpmLN}
satisfy $f_+(x)=f_-(-x)$. Then the spectral action of Theorem \ref{SpActLN} is given by
\begin{equation}\label{fSpAcLNeven}
\Tr(f(D/\Lambda)) = \frac{4}{N} \Lambda^3 \widehat f_+^{(2)}(0) + O(\Lambda^{-k}).
\end{equation}
\end{cor}

\proof The function $\widehat f_\pm^{(k)}$ is the Fourier transform of $v^k f_\pm(v)$, so that
$$ \widehat f_\pm^{(k)}(0) = \int_\R v^k f_{\pm}(v)\, dv = \int_{\R_{\pm}} v^k f_{\pm}(v)\, dv. $$
Using $f_+(v)=f_-(-v)$ one sees that 
$$ \int_0^\infty v^2 f_+(v) dv = \int_{-\infty}^0 v^2 f_-(v) dv $$
so that $\widehat f^{(2)}_+(0)=\widehat f^{(2)}_-(0)$, while
$$ \int_0^\infty v f_+(v) dv = - \int_{-\infty}^0 v f_-(v) dv, $$
so that $\widehat f^{(1)}_+(0)=-\widehat f^{(1)}_-(0)$.  The $\Lambda^2$-terms then
cancel.
\endproof

\subsection{The other spectrum}

We also show, in a similar way, how one can compute the spectral action
for an operator $D$ whose spectrum is of the form given in \eqref{SpLN2spin}.
As before, the multiplicity in row $(i)$ is already in a nice form, while for row $(ii)$ 
we obtain the following.

\begin{lem}\label{spmultLN2spin}
The multiplicity of $-\frac{1}{2} \pm m$ in \eqref{SpLN2spin} is given by 
\begin{equation}\label{multLNspin2}
\begin{array}{lll}
\displaystyle{\frac{2m(m-j)}{N}} & \text{for } \,\,  m\equiv j \mod N, & j=0,1,\ldots, \frac{N}{2}+1 \\[3mm]
\displaystyle{\frac{2m(m-j + N)}{N}} & \text{for } \,\, m\equiv j \mod N, & j=\frac{N}{2}+2, \ldots, N-1.
\end{array}
\end{equation}
\end{lem}

\proof    To handle row $(ii)$, we need to break it up into the pieces $m \equiv j$ (mod $N$), where $j \in \{0,1,\ldots, N-1\}$.  Similar to the previous spectrum, to find nice expressions for the multiplicities, it will be convenient to replace the upper and lower bounds with the highest and lowest values of $iN + \frac{N}{2}$ that satisfy the inequality. 

We first consider the case with $m \equiv j$ (mod $N$), $j = 0,1,\ldots, N/2 + 1$.
The bound $1-m < iN + \frac{N}{2} \leq m-2$ can be replaced by
$$ \frac{N}{2} + j - m \leq iN + \frac{N}{2} \leq m - j - \frac{N}{2} $$
by adding $\frac{N}{2} + j - 1$ to the left hand side and subtracting
$j - 2 + \frac{N}{2}$ from the right hand side.  We check that 
$0 < \frac{N}{2} + j -1 \leq N$, and that $0 \leq j - 2 + \frac{N}{2} < N$.
If $m = kN + j$, where $k= 0,1,2,\ldots$, then we see that there are $2k$ values of $i$ which satisfy the inequality, and hence $-1/2 \pm m$ has multiplicity $2k  m = (2m(m-j))/N$. Once again, when $m$ is $0$ or $1$, this formula gives us a multiplicity of zero, as is necessary, since in row $(ii)$, the index $m$ begins at $m=2$.

We then look at the case with $m \equiv j$ (mod $N$), $j = N/2 + 2, \ldots, N - 1$.
Here the range $1-m < iN + \frac{N}{2} \leq m-2$ becomes
$$ j - m -\frac{N}{2}\leq iN + \frac{N}{2} \leq m - j + \frac{N}{2} $$
by adding $j - 1 - \frac{N}{2}$ to the left hand side and subtracting 
$j - 2 - \frac{N}{2}$ from the right hand side.  Again, we check that $0 < j - 1 - \frac{N}{2} \leq N$, and that $0 \leq j - 2 - \frac{N}{2} < N$.
If $m = kN + j$, where $k= 0,1,2,\ldots$, then we see that there are $2k+2$ values of $i$ which satisfy the inequality, hence $-1/2 \pm m$ has multiplicity $(2k+2) m = (2m(m-j + N))/N$.
\endproof

We can then compute the polynomials that interpolate the spectral multiplicities in the
following way.

\begin{lem}\label{PipmLN2spin}
For $m>0$, the multiplicity of $-1/2+m$, for $m \equiv j$ mod $N$, is given by the values
$P^+_j(-1/2+m)$ of the polynomials
\begin{equation}\label{Pplus2spin}
\begin{array}{rll}
P_j^+ (u)= &   \frac{2}{N} u^2 +  \frac{2-2j}{N} u + \frac{1-2j}{2N} & \text{ for } j = 0, \ldots , \frac{N}{2}+ 1
\\[3mm]
P_j^+ (u)= & \frac{2}{N} u^2 +  \frac{2-2j+2N}{N} u + \frac{1 - 2j + 2N}{2N} & \text{ for } 
j = \frac{N}{2}+ 2,\ldots, N-1.
\end{array}
\end{equation}
For $m\geq 0$, the multiplicity of $-1/2-m$, for $m \equiv \ell$ mod $N$, is given by the
values $P^-_\ell(-1/2-m)$ of the polynomials
\begin{equation}\label{Pminus2spin}
\begin{array}{rll}
P_j^- (u) = & \frac{2}{N} u^2 +  \frac{2+2j}{N} u + \frac{1+2j}{2N}&  \text{ for } 
j = 0, \ldots , \frac{N}{2}+ 1 \\[3mm]
P_j^- (u) = & \frac{2}{N} u^2 + \frac{2+2j-2N}{N} u + \frac{1 + 2j - 2N}{2N} , & \text{ for } 
j = \frac{N}{2}+ 2,\ldots, N-1.
\end{array}
\end{equation}
The multiplicity of $-(N' + iN) - \frac{1}{2}$, for $i\geq 0$, is given by the value $P^-(-(N'+iN)-1/2)$
of the polynomial
\begin{equation}\label{Pminextra2spin}
P^-(u) = -2u -1
\end{equation}
\end{lem}

\proof This follows directly from the expressions for the multiplicities given in 
Lemma \ref{spmultLN2spin} above. 
\endproof

We then have the following result.

\begin{thm}\label{SpAcplusLN2spin}
Let $f_+$ be a Schwartz function supported on the positive reals, chosen as in
Lemma \ref{SpactpmLN}, and let $D$ be an operator with spectrum given by
\eqref{SpLN2spin}. The spectral action
is of the form
\begin{equation}\label{SpAc2spinplus}
\Tr(f_+(D/\Lambda)) = \frac{1}{N} \left( 2 \Lambda^3 \widehat f_+^{(2)}(0)  - \Lambda^2 \widehat f_+^{(1)}(0)  - \Lambda \widehat f_+(0) \right) + O(\Lambda^{-k}) .
\end{equation}
\end{thm}

\proof We set $g_j ^+ (u) = P_j^+(u) f_+(u/ \Lambda)$, with $P^+_j$ as in \eqref{Pplus2spin},
for $j=0,\ldots, N-1$. Then, arguing as in Theorem \ref{posSpAcLN}, we see that the
spectral action is computed by
$$ \Tr(f(D/\Lambda)) = \frac{1}{N} \sum_{j=0}^{N-1} \widehat g_j^+ (0) + O(\Lambda ^{-k}). $$
We can compute each term explicitly using \eqref{Pplus2spin}, and we obtain
\begin{equation}\label{hatg0plus2spin}
\begin{array}{rll}
\widehat g_j^+ (0) =& \displaystyle{\frac{2}{N} \Lambda^3 \widehat f_+^{(2)}(0) +  \frac{2-2j}{N} \Lambda^2 \widehat f_+^{(1)}(0) +  \frac{1-2j}{2N} \Lambda \widehat f_+(0)} \\[2mm]
& \text{ for }  j = 0, \ldots, \frac{N}{2}+1\\[3mm]
\widehat g_j^+ (0) =& \displaystyle{ \frac{2}{N} \Lambda^3 \widehat f_+^{(2)}(0) + \frac{2-2j + 2N}{N} \Lambda^2 \widehat f_+^{(1)}(0) +  \frac{1-2j + 2N}{2N} \Lambda \widehat f_+(0)} , \\[2mm]
& \text{ for } j = \frac{N}{2} + 2 \ldots, N-1. \\
\end{array}
\end{equation}
We have 
$$  \sum_{j=0}^{N/2+1} \frac{2-2j}{N} + \sum_{j=N/2+2}^{N-1} \frac{2-2j + 2N}{N} = -1, $$
$$ \sum_{j=0}^{N/2+1} \frac{1-2j}{2N}  +\sum_{j=N/2+2}^{N-1}  \frac{1-2j + 2N}{2N} = -1. $$
This then gives
$$ \sum_{j=0}^{N-1} \widehat g_j^+ (0)   =  2  \Lambda^3 \widehat f_+^{(2)}(0) - \Lambda^2 \widehat f_+^{(1)}(0) - \Lambda \widehat f_+(0). $$
\endproof

We then proceed in a way similar to Theorem \ref{negSpAcLN} for the case of a test function 
supported on the negative reals.

\begin{thm}\label{SpAcplusLN2spin}
Let $f_-$ be a Schwartz function supported on the negative reals, chosen as in
Lemma \ref{SpactpmLN}. The spectral action for an operator $D$ with 
spectrum given by \eqref{SpLN2spin} is given by
\begin{equation}\label{SpAc2spinminus}
\Tr(f_-(D/\Lambda)) = \frac{1}{N} \left( 2 \Lambda^3 \widehat f_-^{(2)}(0)  + 3\Lambda^2 \widehat f_-^{(1)}(0)  + \Lambda \widehat f_-(0) \right) + O(\Lambda^{-k}).
\end{equation}
\end{thm}

\proof We set  $g_j ^- (u) = P_j^-(u) f_-(u/ \Lambda)$, with $P^-_j$ as in \eqref{Pminus2spin}, and
$g^-(u)=P^-(u) f_-(u/\Lambda)$, for $P^-$ as in \eqref{Pminextra2spin}. By the same reasoning of
Theorem \ref{negSpAcLN} we see that, up to an error term of the order of $O(\Lambda^{-k})$, the
spectral action $\Tr(f_-(D/\Lambda))$ is given by
$$ \frac{1}{N}\left( \widehat g^-(0) + \sum_{j=0}^{n-1} \widehat g^-_j(0) \right) . $$
We then find
\begin{equation}\label{hatg0plus2spin}
\begin{array}{rll}
\widehat g_j^- (0) = & \displaystyle{ \frac{2}{N} \Lambda^3 \widehat f_-^{(2)}(0) +  \frac{2+2j}{N} \Lambda^2 \widehat f_-^{(1)}(0) +  \frac{1+2j}{2N} \Lambda \widehat f_-(0)} , \\[2mm] & 
\text{ for } j = 0, \ldots, \frac{N}{2}+1 \\[3mm]
\widehat g_j^- (0) = & \displaystyle{\frac{2}{N} \Lambda^3 \widehat f_-^{(2)}(0) + \frac{2+ 2j - 2N}{N} \Lambda^2 \widehat f_-^{(1)}(0) + \frac{1 + 2j - 2N}{2N} \Lambda \widehat f_-(0) }\\[2mm] 
& \text{ for } j = \frac{N}{2} + 2, \ldots, N-1
\end{array}
\end{equation}
and
$$ \widehat g^-(0) = -2\Lambda^2 \widehat f_-^{(1)}(0) - \Lambda \widehat f_-(0). $$
We have
$$ \sum_{j=0}^{N/2+1} \frac{2+2j}{N} + \sum_{j=N/2+2}^{N-1} \frac{2+ 2j - 2N}{N} -1 = 3 $$
$$ \sum_{j=0}^{N/2+1}  \frac{1+2j}{2N} + \sum_{j=N/2+2}^{N-1} \frac{1 + 2j - 2N}{2N} -1= 1$$
so we obtain
$$ \widehat g^-(0) + \sum_{j=0}^{n-1} \widehat g^-_j(0) =2 \Lambda^3 \widehat f_-^{(2)}(0) +
3 \Lambda^2 \widehat f_-^{(1)}(0) + \Lambda \widehat f_-(0) . $$
\endproof

We then assemble these two cases together
and we obtain the following expression for the spectral action.

\begin{thm}\label{SpAcpm2spin}
Let $f$ be a Schwartz function on the real line, and let $D$ be an operator with spectrum given by
\eqref{SpLN2spin}. For $f_+$ and $f_-$ chosen
as in Lemma \ref{SpactpmLN}, with $f=f_+ + f_-$ on an open neighborhood of the
spectrum of $D$, the spectral action is given by
\begin{equation}\label{SpActL2N2spin}
\begin{array}{rl}
\Tr(f(D/\Lambda)) = & \displaystyle{ \frac{1}{N} } \left(
2 \Lambda^3 (\widehat f_+^{(2)}(0) + \widehat f_-^{(2)}(0))  \right. \\[2mm] 
 + & \left. \Lambda^2 (3 \widehat f_-^{(1)}(0) -
\widehat f_+^{(1)}(0)  ) + \Lambda ( \widehat f_-(0) -  \widehat f_+(0) )\right)
\end{array}
\end{equation}
up to an error term of the order $O(\Lambda^{-k})$ for arbitary $k>0$.
\end{thm}

In particular, if the function $f$ is an even function,
then $\widehat f_+^{(2)}(0)=\widehat f_-^{(2)}(0)$ and $\widehat f_-(0) = \widehat f_+(0)$,
while $\widehat f_-^{(1)}(0)=-\widehat f_+^{(1)}(0)$, so one obtains
$$ \Tr(f(D/\Lambda)) = 4 \Lambda^3 \widehat f_+^{(2)}(0) - 4 \Lambda^2 \widehat f_+^{(1)}(0). $$

We see then that the resulting spectral action for the two spectra \eqref{lensspectrum}
and \eqref{SpLN2spin} of \cite{Bar2} is {\em different}, 
unlike what we have seen in all the other explicit
cases of Dirac spectra on manifolds for which we explicitly computed the spectral action,
where the spectral action is independent of the spin structure, even though the Dirac
spectrum itself may be different for different spin structures. This is in clear contrast with the
lens spaces calculation with the correct Dirac spectrum that we describe in Section \ref{LensTrueSec}, below.

However, it is interesting to notice that the two spectra \eqref{lensspectrum}
and \eqref{SpLN2spin} have the property that the spectral action computed 
for the operator $|D|$ instead of $D$ restores the symmetry, namely it gives
the same result for the two spectra.

\subsection{The spectral action for $|D|$}\label{LNspsignSec}

We consider again an operator $D$ that has as spectrum either \eqref{lensspectrum}
or \eqref{SpLN2spin}. We replace $D$ by $|D|$ and we proceed to the same
calculation of the spectral action as before.

\begin{thm}\label{SpAcLNsign}
Let $D=F|D|$ be an operator with spectrum \eqref{lensspectrum}.
Let $f$ be an even Schwartz function and $f_+$ and $f_-$ be as in Lemma \ref{SpactpmLN},
with $f=f_+ + f_-$ on an open neighborhood of the spectrum of $D$, and with $f_-(-x)=f_+(x)$.
Then the spectral action $\Tr(f(|D|/\Lambda))$ is given by 
\begin{equation}\label{SpAcsignDLN}
\Tr(f(|D|/\Lambda)) = \frac{1}{N}\left( 4  \Lambda^3 \widehat f^{(2)}_+(0) + 2 \Lambda^2 \widehat
 f^{(1)}_+(0) \right) +O(\Lambda^{-k}).
\end{equation} 
\end{thm}

\proof
Let $\lambda^\pm_{j,m}=\pm m-1/2$ and $\lambda^-_i=-iN-1/2$ be the arithmetic
progressions of the Dirac spectrum \eqref{lensspectrum} on $\cL_N$ with
the canonical spin structure. We have obtained in Lemma \ref{posmultLN}
polynomials $P^+_j(u)$, $P^-_j(u)$ and $P^-(u)$ such that,
for $m>0$, $P^+_j(\lambda^+_{j,m})$ is the spectral multiplicity of $\lambda^+_{j,m}$,
while for $m\leq 0$, $P^-_j(\lambda^-_{j,m})$ is the spectral multiplicity of $\lambda_{j,m}^-$,
and $P^-(\lambda^-_i)$ is the spectral multiplicity of $\lambda^-_i$.

When we replace $D$ by $|D|$, we want new polynomials
$\bar P^-_j(u)$ and $\bar P^-(u)$, with the property that, for $m\leq 0$,
$\bar P^-_j(-\lambda_{j,m}^-)$ 
is the spectral multiplicity of  $\lambda_{j,m}^{-}$
and $\bar P^-(-\lambda_i^{-})$ is the spectral multiplicity of $\lambda_i^{-}$.
It suffices to choose $\bar P^-_j(-u)=P^-_j(u)$ and $\bar P^-(-u)=P^-(u)$.
All the polynomials $P^+_j$, $P^-_j$ and $P^-$ of Lemma \ref{posmultLN}
are of the form $c_2 u^2 + c_1 u +c_0$ for suitable coefficients $c_k$. Thus,
while the $P^+_j$ remain the same, the corresponding $\bar P^-_j$
and $\bar P^-$ will be of the form $c_2 u^2 - c_1 u +c_0$. More precisely, we
obtain
\begin{equation}\label{barPiminusLN}
\begin{array}{rl}
\bar P_0^- (u) =& \frac{2}{N}u^2- \frac{2}{N}u+\frac{1}{2N} \\[2mm]
\bar P_1^- (u) =& \frac{2}{N}u^2- \frac{4}{N}u+\frac{3}{2N}\\[2mm]
\bar P_j^- (u) =& \frac{2}{N}u^2 - \frac{2+ 2j - N}{N}u+\frac{1 + 2j - N}{2N}, \quad j = 2,3,\ldots, N-1,
\end{array}
\end{equation}
\begin{equation}\label{barPminusLN}
\bar P^-(u) =2u -1.
\end{equation}
Similarly, for $f$ even with $f=f_+ + f_-$ on an open neighborhood of the spectrum, as before, 
and with $f_+(-u)=f_-(u)$, we have $f_+(-\lambda_{j,m}^-)=f_-(\lambda_{j,m}^-)$ and
$f_+(-\lambda^-_j) = f_-(\lambda^-_j)$. 

Correspondingly, we now set 
\begin{equation}\label{bargminus}
 g_j^+(u) = P^+_j(u) f_+(\frac{u}{\Lambda}), \ \ \  \
\bar g^-_j(u) = \bar P^-_j(u) f_-(\frac{u}{\Lambda}), \ \ \ \  \bar g^-(u) = \bar P^-(u) f_-(\frac{u}{\Lambda}) , 
\end{equation}
We see then that $\Tr(f(|D|/\Lambda))$ is given by
\begin{equation}\label{Spbarg}
\sum_{k\in \Z} \left( \sum_{j=0}^{N-1} g^+_j(kN+\frac{2j-1}{2}) +
\sum_{j=0}^{N-1} \bar g^-_j(kN + \frac{2j+1}{2}) + \bar g^-(kN +\frac{1}{2}) \right).
\end{equation}

We then use Poisson summation as before and we find 
\begin{equation}\label{Spbarhatg}
\Tr(f(|D|/\Lambda)) = \frac{1}{N} \left( \sum_{j=0}^{N-1} \widehat g^+_j(0) +
\sum_{j=0}^{N-1} \widehat {\bar g^-_j } (0) + \widehat { \bar g^- } (0) \right) + O(\Lambda^{-k}).
\end{equation}

We then see that
$$ \sum_{j=0}^{N-1}  \widehat g^+_j(0) = 2 \Lambda^3 \widehat f^{(2)}_+(0) + \Lambda^2 \widehat
 f^{(1)}_+(0) $$
as before, while
$$ \sum_{j=0}^{N-1} \widehat {\bar g^-_j } (0)  + \widehat { \bar g^- } (0) =
2 \Lambda^3 \widehat f^{(2)}_-(0) - \Lambda^2 \widehat f^{(1)}_- (0) =
2 \Lambda^3 \widehat f^{(2)}_+(0) + \Lambda^2 \widehat
 f^{(1)}_+(0) ,$$
so that \eqref{SpAcsignDLN} holds.
\endproof

We now see that this is the same result obtained from the second spectrum \eqref{SpLN2spin}.

\begin{thm}\label{SpAcLNsign2sp}
Let $D=F|D|$ be an operator with spectrum \eqref{SpLN2spin}.
Let $f$ be an even Schwartz function and $f_+$ and $f_-$ be as in Lemma \ref{SpactpmLN},
with $f=f_+ + f_-$ on an open neighborhood of the spectrum of $D$, and with $f_-(-x)=f_+(x)$.
Then the spectral action $\Tr(f(|D|/\Lambda))$ is given by 
\begin{equation}\label{SpAcsignDLN2sp}
\Tr(f(|D|/\Lambda)) = \frac{1}{N}\left( 4  \Lambda^3 \widehat f^{(2)}_+(0) + 2 \Lambda^2 \widehat
 f^{(1)}_+(0) \right) + O(\Lambda^{-k}).
\end{equation} 
\end{thm}

\proof The argument is the same as in the previous case, but applied to the eigenvalues
and multiplicities \eqref{SpLN2spin} and the polynomials $P^+_j$, $P^-_j$ and $P^-$ of
Lemma \ref{PipmLN2spin}. We then compute $\Tr(f(|D|/\Lambda))$ as in the case of
the canonical spin structure, using the corresponding polynomials $\bar P^-_j$ and
$\bar P^-$ and the functions $\bar g^-_j$ and $\bar g^-$ as above. We obtain again
the expression \eqref{Spbarhatg}, where in this case
$$ \sum_{j=0}^{N-1}  \widehat g^+_j(0) = 2 \Lambda^3 \widehat f^{(2)}_+(0) -
\Lambda^2 \widehat f^{(1)}_+(0) - \Lambda \widehat f_+(0) $$
and
$$ \begin{array}{rl}
\displaystyle{ \sum_{j=0}^{N-1} \widehat {\bar g^-_j } (0)  + \widehat { \bar g^- } (0) } = &
\displaystyle{
2 \Lambda^3 \widehat f^{(2)}_- (0) - 3 \Lambda^2 \widehat f^{(1)}_-(0) + \Lambda \widehat f_-(0) }\\[3mm] = & \displaystyle{
2 \Lambda^3 \widehat f^{(2)}_+ (0) + 3 \Lambda^2 \widehat f^{(1)}_+ (0) + \Lambda \widehat f_+(0), }
\end{array} $$
so that $\Tr(f(|D|/\Lambda))$ is given by
$$ \frac{1}{N}\left( 2 \Lambda^3 \widehat f^{(2)}_+(0) -
\Lambda^2 \widehat f^{(1)}_+(0) - \Lambda \widehat f_+(0) +
2 \Lambda^3 \widehat f^{(2)}_+ (0) + 3 \Lambda^2 \widehat f^{(1)}_+ (0) + \Lambda \widehat f_+(0) \right), $$
up to an error term of the order of $O(\Lambda^{-k})$. This
gives again the same \eqref{SpAcsignDLN2sp} as for the canonical spin structure.
\endproof

\subsection{The slow-roll potential: a false positive}

Now we use this result to compute the spectral action for an operator $\cD^2$ with 
\begin{equation}\label{cDS1}
 \cD =\left(\begin{array}{cc}  0 & D \otimes  1 + i \otimes D_{S^1} \\
D \otimes 1 - i \otimes D_{S^1} & 0     \end{array}\right),
\end{equation}
where $D_{S^1}$ has spectrum $\beta^{-1} (\Z + 1/2)$, and $D$ is an operator with
spectrum either \eqref{lensspectrum}
or \eqref{SpLN2spin}. 

The spectrum of the operator $\cD^2$ will be contained in the set of values of the form 
$(\lambda_{j,m}^\pm)^2  (\Lambda a)^{-2}+\lambda_n^2 (\Lambda \beta)^{-2}$ 
and $(\lambda_i^-)^2  (\Lambda a)^{-2} + \lambda_n^2 (\Lambda \beta)^{-2}$,
where $\lambda_{j,m}^\pm$ and $\lambda_i^-$ are the arithmetic
progressions associated to the spectrum of  $D$ 
and $\lambda_n = n+1/2$ are the eigenvalues on a circle of radius one.
We see that the pairs of points $(u,v)$ in $\R^2$ which are of the form
$(\lambda_{j,m}^\pm,\lambda_n)$ or $(\lambda_i^-, \lambda_n)$ all
lie outside of a vertical strip around $u=0$. We fix a value $\alpha <1$ such
that the strip $u \in (-\alpha,\alpha)$ contains on such pair.  We also fix
a $k >0$, which will determine the order $O(\Lambda^{-k})$ of error
in the spectral action computation. Let then $\ell(u^k)$ be a smooth
function, which is equal to zero for $u \leq 0$ and is equal to one for
$u \geq \alpha$.

\begin{lem}\label{estimateell}
Suppose given a polynomial $P(x)=c_2 x^2 + c_1 x + c_0$, and let $h$ be a
Schwartz function on $\R$. The difference between the integrals
\begin{equation}\label{intellh}
I_1 = \int_{\R^2} P(x)\, \ell(x^k (\Lambda a)^k)\, h(x^2 + y^2) \, dx \, dy 
\end{equation}
and
\begin{equation}\label{inth}
I_2 = \int_0^\infty \int_{-\pi/2}^{\pi/2} (\frac{c_2}{2} \rho^2 + c_1 \rho \cos\theta + c_2) \, h(\rho^2) \,\rho\, 
d\rho \, d\theta 
\end{equation}
is bounded by
\begin{equation}\label{I12bound}
| I_1 - I_2 | = O(\Lambda^{-k}) .
\end{equation}
\end{lem}

\proof The difference $I_1 - I_2$ is computed by
$$ \left| \int_\R \left( \int_0^\alpha P(x)\, \ell(x^k (\Lambda a)^k)\, h(x^2 + y^2) \, dx \right) \, dy \right| 
\leq C \frac{\alpha^k}{(\Lambda a)^k}. $$
\endproof

We can then proceed to compute the spectral action.

\begin{thm}\label{SpActLNS1}
The spectral action for the operator $\cD$ of \eqref{cDS1}, where
$D$ has spectrum \eqref{lensspectrum} is
of the form
\begin{equation}\label{SpALNS1}
\Tr(h(\cD^2/\Lambda^2)) =  2\pi \Lambda^4 a^3 \beta \int_0^\infty u\, h(u)\, du +
2\Lambda^3 a^2 \beta \int_0^\infty u^{1/2} h(u) \, du
+ O(\Lambda^{-k}).
\end{equation}
\end{thm}

\proof For a given $k>0$, we choose a cutoff $\ell(u^k)$ as in Lemma \ref{estimateell}. 
We then set $\ell_+(u) = \ell(u)$ and $\ell_-(u) = \ell_+(-u)$.
We then consider, for $j=0,\ldots N-1$, functions of the form
\begin{equation}\label{gjplusuv}
g_j^+(u,v)= 2P_j^+(u)\, \ell_+(u^k) \, h(u^2 (\Lambda a)^{-2} + v^2 (\Lambda \beta)^{-2}),
\end{equation}
where the polynomials $P_j^+(u)$ are as in \eqref{PiplusLN}. We also set
\begin{equation}\label{gjminusuv}
g_j^-(u,v)= 2 \bar P_j^-(u)\, \ell_-(u^k)\, h(u^2 (\Lambda a)^{-2} + v^2 (\Lambda \beta)^{-2}),
\end{equation}
with the polynomials $\bar P^-_j(u)$ of \eqref{barPiminusLN} and
\begin{equation}\label{gminusuv}
g^-(u,v)= 2\bar P^-(u)\, \ell_-(u^k)\, h(u^2 (\Lambda a)^{-2} + v^2 (\Lambda \beta)^{-2}),
\end{equation}
with $\bar P^-(u)$ as in \eqref{barPminusLN}.
The spectral action $\Tr(h(\cD^2/\Lambda^2))$, for $D$ with spectrum \eqref{lensspectrum},
is then given by 
\begin{equation}\label{SpAchgjpmLN}
\begin{array}{rl}
\Tr(h(\cD^2/\Lambda^2)) = & \displaystyle{ \sum_{j=0}^{N-1} g^+_j(nN +\frac{(2j-1)}{2}, m+\frac{1}{2}) } 
\\[3mm]
+ &  \displaystyle{ \sum_{j=0}^{N-1} g^-_j(nN - \frac{(2j+1)}{2}, m+\frac{1}{2}) } \\[3mm]
+ &   \displaystyle{ g^-(nN -\frac{1}{2}, m+ \frac{1}{2}) }.
\end{array}
\end{equation}
We compute it by applying the Poisson summation formula to the functions
\eqref{gjplusuv}, \eqref{gjminusuv}, and \eqref{gminusuv}. We obtain, as in
the previous cases
\begin{equation}\label{SpAcLNS1g00}
\Tr(h(\cD^2/\Lambda^2)) = \frac{1}{N}\left( \sum_{j=0}^{N-1}
\widehat g_j^+(0,0) + \sum_{j=0}^{N-1} \widehat g_j^-(0,0) + \widehat 
g^-(0,0) \right) + O(\Lambda^{-k}).
\end{equation}
We then use Lemma \ref{estimateell} to estimate the integrals, up to
an error term of order $O(\Lambda^{-k})$, to be of the form 
\begin{equation}\label{SpLNS1plus}
\begin{array}{rl} \displaystyle{
\sum_{j=0}^{N-1} \widehat g_j^+(0,0)} = &  \displaystyle{
2 \Lambda^2 a \beta \int_0^\infty \int_{-\pi/2}^{\pi/2} \left( \rho^2 (\Lambda a)^2 + \rho \cos\theta (\Lambda a) \right) \, h(\rho^2) \, \rho\, d\rho\, d\theta} \\[3mm]  = &  \displaystyle{ 2\pi \Lambda^4 a^3 \beta \int_0^\infty \rho^3 h(\rho^2) d\rho
+ 2 \Lambda^3 a^2 \beta \int_0^\infty \rho^2 h(\rho^2) d\rho} ,
\end{array}
\end{equation}
where we used the fact that 
$$ \sum_{j=0}^{N-1} P^+_j(u) = 2 u^2 +u. $$
After a change of variables, we write the above as
$$ \sum_{j=0}^{N-1} \widehat g_j^+(0,0) = \pi \Lambda^4 a^3 \beta \int_0^\infty u h(u) \, du +
\Lambda^3 a^2 \beta \int_0^\infty u^{1/2} h(u) \, du. $$
Similarly, using the approximation of Lemma \ref{estimateell} we obtain, up to
an error term of the order of $O(\Lambda^{-k})$,
\begin{equation}\label{SpLNS1minus}
\begin{array}{rl} \displaystyle{
\sum_{j=0}^{N-1} \widehat g_j^-(0,0) + \widehat g^-(0,0) }
= &  \displaystyle{ 2\pi \Lambda^4 a^3 \beta \int_0^\infty \rho^3 h(\rho^2) d\rho
+ 2 \Lambda^3 a^2 \beta \int_0^\infty \rho^2 h(\rho^2) d\rho} \\[3mm]
= & \displaystyle{ \pi \Lambda^4 a^3 \beta \int_0^\infty u h(u) \, du +
\Lambda^3 a^2 \beta \int_0^\infty u^{1/2} h(u) \, du },
\end{array}
\end{equation}
where we used the fact that 
$$ \sum_{j=0}^{N-1} \bar P^-_j(u) + \bar P^-(u) = 2 u^2 -u $$
and that $\ell_-(u)=\ell_+(-u)$. This then gives \eqref{SpALNS1}.
\endproof

The case where $D$ has spectrum \eqref{SpLN2spin} is analogous and
yields the same result.

\begin{thm}\label{SpActLNS12sp}
Consider the operator $\cD$ of \eqref{cDS1}, where $D$ has spectrum
\eqref{SpLN2spin}. The spectral action is
of the form
\begin{equation}\label{SpALNS1spin2}
\Tr(h(\cD^2/\Lambda^2)) =  2\pi \Lambda^4 a^3 \beta \int_0^\infty u\, h(u)\, du +
2\Lambda^3 a^2 \beta \int_0^\infty u^{1/2} h(u) \, du
+ O(\Lambda^{-k}).
\end{equation}
\end{thm}

\proof One proceeds exactly as in Theorem \ref{SpActLNS1}, but using the
expressions for $P^+_j$, $\bar P^-_j$ and $\bar P^-$ as in Theorem \ref{SpAcLNsign2sp}.
One then has 
$$ \sum_{j=0}^{N-1} P^+_j(u) = 2 u^2 -u -1, \ \ \ \ 
 \sum_{j=0}^{N-1} \bar P^-_j(u) + P^-(u) = 2 u^2 - 3 u +1,
$$
so that, using $\ell_-(u)=\ell_+(-u)$, one correspondingly obtains
\begin{equation}\label{SpLNS1plus2sp}
\begin{array}{rl} \displaystyle{
\sum_{j=0}^{N-1} \widehat g_j^+(0,0)} = &  \displaystyle{
2 \pi \Lambda^4 a^3 \beta \int_0^\infty \rho^3 h(\rho^2) d\rho 
- 2 \Lambda^3 a^2 \beta \int_0^\infty \rho^2 h(\rho^2) d\rho } \\[3mm] - & \displaystyle{ 2\pi \Lambda^2 a\beta 
\int_0^\infty h(\rho^2) \rho d\rho} \\[3mm]
=& \displaystyle{ \pi \Lambda^4 a^3 \beta \int_0^\infty u \, h(u) \,du 
- \Lambda^3 a^2 \beta \int_0^\infty u^{1/2} h(u) \, du } \\[3mm] 
- & \displaystyle{ \pi \Lambda^2 a \beta \int_0^\infty h(u) du. }\end{array}
\end{equation}
Similarly, one obtains
\begin{equation}\label{SpLNS1minus2sp}
\begin{array}{rl} \displaystyle{
\sum_{j=0}^{N-1} \widehat g_j^-(0,0) + \widehat g^-(0,0) }
= &  \displaystyle{   \pi \Lambda^4 a^3 \beta \int_0^\infty u \, h(u) \, du } \\[3mm]
+ & \displaystyle{ 3 \Lambda^3 a^2 \beta \int_0^\infty u^{1/2} h(u) du } \\[3mm]
+ & \displaystyle{\pi \Lambda^2 a \beta \int_0^\infty h(u) du.}
\end{array}
\end{equation}
Thus, adding these contributions one obtains then the same \eqref{SpALNS1spin2}
as in the previous case. 
\endproof 

We then see that the form of the associated potential $V(\phi)$ coming from
the perturbations $\cD^2 \mapsto \cD^2 + \phi^2$ is very different from the
3-sphere and the other spherical manifolds computed in this paper, quaternionic
and dodecahedral space.

We set
\begin{equation}\label{cVcZ}
\cV(x) =\int_0^\infty u\, (h(u+x) - h(u)) \, du , \ \ \  \cZ(x) = \int_0^\infty u^{1/2} \, (h(u+x) - h(u)) \, du,
\end{equation}
in the variable $x=\phi^2/\Lambda^2$.

\begin{prop}\label{PertSALNS1}
Let $\cD$ be the operator of \eqref{cDS1}. 
We have
\begin{equation}\label{hVZ}
\Tr(h((\cD^2+\phi^2)/\Lambda^2)) = \Tr(h(\cD^2/\Lambda^2)) + 2 \pi \Lambda^4 a^3 \beta \cV(\phi^2/\Lambda^2) + 2 \Lambda^3 a^2 \beta \cZ(\phi^2/\Lambda^2).
\end{equation}
Thus, the potential $V(x)$, for $x=\phi^2/\Lambda^2$ is of the form
\begin{equation}\label{Vphilens}
V(x)=  2 \pi \Lambda^4 a^3 \beta\, \cV(\phi^2/\Lambda^2) + 2 \Lambda^3 a^2 \beta \,\cZ(\phi^2/\Lambda^2).
\end{equation}
\end{prop}

\proof This is an immediate consequence of Theorems \ref{SpActLNS1}
and \ref{SpActLNS12sp}. \endproof

We then see that the form of the slow-roll parameters is also different in this
case.

\begin{prop}\label{slowrollLN}
The slow-roll parameters for the potential $V(x)$ are given by
\begin{equation}\label{epsilonetaLN}
\epsilon(x) = \frac{m_{Pl}^2}{8\pi} A, \ \ \ \text{ and } \ \ \ 
\eta(x) = \frac{m_{Pl}^2}{8\pi} (B-A), 
\end{equation}
where
\begin{equation}\label{AtermLN}
A=\frac{1}{2} \left(\frac{ \pi C \cV^\prime (x) + \cZ^\prime(x) }{\pi C \cV(x) + \cZ(x)} \right)^2
\end{equation}
\begin{equation}\label{BtermLN}
B = \frac{\pi C \cV^{\prime\prime}(x) + \cZ^{\prime\prime}(x)}{\pi C \cV(x) + \cZ(x)} .
\end{equation}
\end{prop}

\proof This follows directly from the definition of the slow-roll parameters,
having imposed the condition $\Lambda(t)\sim 1/a(t)$, so that $\Lambda a=C$, 
on the Friedmann form of the spacetime back in Lorentzian signature. \endproof

This calculation creates a ``false positive" which gives the impression that
there are spherical manifolds for which the inflation potential and slow-roll
parameters are genuinely different from those of the sphere. This would
make for a much stronger correlation between inflation and cosmic topology
than what we have observed in the previous section, with different inflation
scenarios not only between spherical and flat cases, but even between 
different topologies with the same underlying spherical geometry. However,
this turns out not to be the case. The true story of the lens spaces, described
in the coming section, shows that in fact, with the correct calculation of the
Dirac spectrum, they behave exactly as the other spherical topology, with
the same slow-roll parameters as in the simply connected case.

\medskip

\subsection{Lens spaces: a discrepancy}\label{LensTrueSec}

In this section we compute the Dirac spectrum for lens space using the same generating function technique due to B\"ar \cite{Bar}, that we used to compute the Dirac spectrum for the Poincar\'e homology sphere, and compare the results to the calculation in \cite{Bar2}.

For simplicity, let us just consider the space $\cL _N = SU(2)/ \Z _N$ in the case $N =4$ with the canonical spin structure. By applying equations \eqref{Spmultgen}, \eqref{Spmultgen2}, one obtains that the generating functions for the spectral multiplicities for $\cL _4$, with the canonical spin structure, are given by:

\begin{align} \label{lens4plus}
F_+ (z) &= - \frac{2(z + 5z^3 + z^5 + z^7)}{(-1 + z^2)^3 (1+z^2)^2},
\\ \label{lens4minus}
F_-(z) &= - \frac{2(1 + z^2 + 5z^4 + z^6)}{(-1 + z^2)^3 (1+z^2)^2}.
\end{align}

Proceeding in precisely the same manner as in the case of the Poincar\'e homology sphere, one obtains the following lemma.

\begin{lem}\label{lens4polys}
There are polynomials $P_k (u)$, for $k = 0, \ldots , 3$, so that $P_k(3/2 + k + 4j) = m(3/2 + k + 4j, D)$ for all $j \in \Z$.  The $P_k(u)$ are given as follows:

\begin{align*}
P_k &= 0, \quad \mathrm{whenever~} k \mathrm{~is~even}. \\
P_1 (u) &= \frac{1}{8} - \frac{1}{2}u + \frac{1}{2}u^2. \\
P_3 (u) &= -\frac{3}{8} + \frac{1}{2}u + \frac{1}{2}u^2.
\end{align*}

\end{lem}

Before comparing these multiplicities with the ones given in \cite{Bar2}, let us first compute the nonperturbative spectral action of the lens space.

\begin{thm}\label{SpActLens4}
Let $D$ be the Dirac operator on $\cL _4$, with the canonical spin structure. Then, for $f$ a Schwartz function, the spectral action is
given by
\begin{equation}\label{SALens}
\Tr(f(D/\Lambda)) = \frac{1}{4} \left(  \Lambda^3 \widehat f^{(2)}(0) -\frac{1}{4}\Lambda \widehat f(0)  \right),
\end{equation}
which  is precisely 1/4 of the spectral action on the sphere.
\end{thm}

\proof As usual the result follows by applying Poisson summation to the
functions $g_j(u)=P_j(u) f(u/\Lambda)$. This
gives, up to an error term which is of the order of $O(\Lambda^{-k})$ for
any $k>0$, the spectral action in the form 
$$ \Tr(f(D/\Lambda)) = \frac{1}{4} \sum_{j=0}^{3} \widehat g_j (0)= 
\frac{1}{4} \int_\R \sum_j P_j(u) f(u/\Lambda) du. $$ 
It suffices then to notice that
$$ \sum_{j=0}^{3} P_j(u) =  u^2-\frac{1}{4}. $$
The result then follows as in the sphere case.
\endproof

Observe that this time around, the spectral action is a constant multiple of the spectral action of the sphere, and so one obtains the same slow-roll parameters as in the simply connected case, just as with the other spherical space forms.

Let us compare the multiplicities obtained using the generation function method in Lemma \ref{lens4polys} with those obtained using the results of \cite{Bar2} in Lemma \ref{posmultLN}.  By setting $N =4$ in Lemma \ref{posmultLN} it is immediately evident that the two sets of multiplicities do not agree.

As a side remark, even if we replace $-(m+1) < iN$ with $-(m-2) \leq iN$ in equation (\ref{lensspectrum}) when performing the computation of Lemma \ref{posmultLN}, this just results in altering $P^{\pm}_0$, and $P^{\pm}_1$ very slightly, while leaving the other $P^{\pm}_j$ unchanged, and the resulting multiplicities still do not agree with the multiplicities of Lemma \ref{lens4polys}.

We are inclined to believe that the generating function method of Lemma \ref{lens4polys} gives the correct answer because of two reasons.  First, the generating function method leads to a spectral action of $1/|G|$ times the spectral action of $S^3$ where $G$ is the group acting on $S^3$, and this is exactly the result we obtained for the other spherical space forms.  Secondly, if one computes the Dirac spectrum of $SU(2)/Q8$ using the generating function method, one gets the same answer as obtained by Ginoux in \cite{Gin}, where the Dirac spectrum of $SU(2)/Q8$ is computed using representation theoretic methods.


\begin{thebibliography}{99}

\bibitem{ALST} R.~Aurich, S.~Lustig, F.~Steiner, H.~Then, {\em Cosmic
microwave background alignment in multi-connected universes}, Class. Quantum Grav.
24 (2007) 1879--1894.

\bibitem{Bar} C.~B\"ar, {\em The Dirac operator on space forms of positive
curvature}, J. Math. Soc. Japan, 48 (1996) N.1, 69--83.

\bibitem{Bar2} C.~B\"ar, {\em The Dirac operator on homogeneous spaces and its spectrum
on 3-dimensional lens spaces},  Arch. Math. Vol.59 (1992) 65--79.

\bibitem{Bar3}  C.~B\"ar, {\em Dependence of Dirac Spectrum on the Spin Structure}, S\'eminaires \& Congr\`es, 4, 2000, 17--33.

\bibitem{dBL} P.~de Bernardis, P.A.R.~Ade, J.J.~Bock, J.R.~Bond, J.~Borrill, A.~Boscaleri, K.~Coble, B.P.~Crill, G.De~Gasperis, P.C.~Farese, P.G.~Ferreira, K.~Ganga, M.~Giacometti, E.~Hivon, V.V.~Hristov, A.~Iacoangeli, A.H.~Jaffe, A.E.~Lange, L.~Martinis, S.~Masi, P.V.~Mason, P.D.~Mauskopf, A.~Melchiorri, L.~Miglio, T.~Montroy, C.B.~Netterfield, E.~Pascale, F.~Piacentini, D.~Pogosyan, S.~Prunet, S.~Rao, G.~Romeo, J.E.~Ruhl, F.~Scaramuzzi, D.~Sforna, N.~Vittorio,
{\em A flat Universe from high-resolution maps of the cosmic microwave background radiation},
Nature 404 (2000), 955--959.

\bibitem{BroSu} T.~van den Broek, W.D.~van Suijlekom, {\em 
Supersymmetric QCD and noncommutative geometry}, arXiv:1003.3788.

\bibitem{CLLLRW} S.~Caillerie, M.~Lachi\`eze-Rey, J.P.~Luminet,
R.~Lehoucq, A.~Riazuelo, J.~Weeks, {\em A new analysis of the
Poincar\'e dodecahedral space model}, Astronomy and Astrophysics, Vol.476
(2007) N.2, 691--696.

\bibitem{ChCo}  A.~Chamseddine, A.~Connes, 
{\em The spectral action principle}. 
Comm. Math. Phys. 186 (1997), no. 3, 731--750.

\bibitem{uncanny} A.~Chamseddine, A.~Connes, 
{\em The uncanny precision of the spectral action},
Commun. Math. Phys. 293 (2010) 867--897.

\bibitem{CCM} A.~Chamseddine, A.~Connes, M.~Marcolli, {\em Gravity and
the standard model with neutrino mixing}, Adv. Theor. Math. Phys.
 11  (2007),  no. 6, 991--1089.
 
 \bibitem{CoSM} A.~Connes, {\it Gravity coupled with matter and
foundation of  noncommutative geometry}. Commun. Math. Phys.,
182 (1996) 155--176.

\bibitem{CSSK} N.J.~Cornish, D.N.~Spergel, G.D.~Starkman, E.~Komatsu,
{\em Constraining the topology of the universe}, Phys. Rev. Lett. 
92 (2004) 201302 [4 pages].

\bibitem{Dahl} M.~Dahl, {\em Prescribing eigenvalues of the Dirac operator}, Manuscripta
Math. 118 (2005) 191--199.

\bibitem{Dahl2} M.~Dahl, {\em Dirac eigenvalues for generic metrics on three-manifolds},
Annals of Global Analysis and Geometry, 24 (2003) 95--100.

\bibitem{dSHW} A.~De Simone, M.P.~Hertzberg, F.~Wilczek, {\em
Running inflation in the Standard Model}, arXiv:0812.4946v2.

\bibitem{GLLUW} E.~Gausmann, R.~Lehoucq, J.P.~Luminet, J.P.~Uzan,
J.~Weeks, {\em Topological lensing in spherical spaces},
Class. Quantum Grav. 18 (2001) 5155--5186. 

\bibitem{Gin} N.~Ginoux, {\em The spectrum of the Dirac operator on $\rm SU\sb 2/Q\sb 8$}.  Manuscripta Math.  125  (2008),  no. 3, 383--409.

\bibitem{GoReTa} G.I.~Gomero, M.J.~Reboucas, R.~Tavakol, {\em
Detectability of cosmic topology in almost flat universes},
Class. Quant. Grav. 18 (2001) 4461--4476.

\bibitem{GoReTe} G.I.~Gomero, M.J.~Reboucas, A.F.F.~Teixeira, {\em Spikes in
cosmic crystallography II: topological signature of compact flat universes}, Phys.
Lett. A 275 (2000) 355--367.

\bibitem{Hitchin} N.~Hitchin, {\em Harmonic spinors}, Advances Math. 14 (1974) 1--55.

\bibitem{KaSpeSu} M.~Kamionkowski, D.N.~Spergel, N.~Sugiyama, {\em Small-scale
cosmic microwave background anisotropies as a probe of the geometry of the universe},
Astrophysical J. 426 (1994) L 57--60.

\bibitem{LaLu} M.~Lachi\`eze-Rey, J.P.~Luminet, {\em Cosmic topology.} Physics Reports,
254 (1995) 135--214.

\bibitem{LWUGL} R.~Lehoucq, J.~Weeks, J.P.~Uzan, E.~Gausmann, J.P.~Luminet,
{\em Eigenmodes of three-dimensional spherical spaces and their applications to cosmology.}
Class. Quantum Grav. 19 (2002) 4683--4708.

\bibitem{LWRL} J.P.~Luminet,  J.~Weeks, A.~Riazuelo, R.~Lehoucq, {\em Dodecahedral space topology as an explanation for weak wide-angle temperature correlations in the cosmic microwave 
background},  Nature 425 (2003) 593--595.

\bibitem{MaPie} M.~Marcolli, E.~Pierpaoli, {\em Early universe models from noncommutative
geometry},  arXiv:0908.3683.

\bibitem{McInnes} B.~McInnes, {\em APS instability and the topology of the brane-world.}
Physics Letters B, Vol.593 (2004) N.1-4, 10--16.

\bibitem{NeSa} W.~Nelson, M.~Sakellariadou, 
{\em Natural inflation mechanism in asymptotic noncommutative geometry}, 
Phys. Lett. B (2009) Vol.680, 263--266.

\bibitem{NiarJaffe} A.~Niarchou, A.~Jaffe, {\em Imprints of spherical
nontrivial topologies on the cosmic microwave background}, Physical
Review Letters, 99 (2007) 081302, 4pp. 

\bibitem{OCTZH} A.~de Oliveira-Costa, M.~Tegmark, M.~Zaldarriaga, A.~Hamilton,
{\em Significance of the largest scale CMB fluctuations in WMAP}, Phys. Rev. D 69
(2004) 063516 [12 pages].

\bibitem{Pfa} F.~Pf\"affle, {\em The Dirac spectrum of Bieberbach manifolds}, J. Geom. Phys.
35 (2000) 367--385.

\bibitem{RULW} A.~Riazuelo, J.P.~Uzan, R.~Lehoucq, J.~Weeks, 
{\em Simulating Cosmic Microwave Background maps in multi-connected spaces},
Phys.Rev. D69 (2004) 103514 [28 pages].

\bibitem{RWULL} A.~Riazuelo, J.~Weeks, J.P.~Uzan, R.~Lehoucq, J.P.~Luminet, 
{\em Cosmic microwave background anisotropies in multiconnected flat spaces},
Phys. Rev. D 69 (2004) 103518 [25 pages].

\bibitem{RouRo} B.F.~Roukema, P.T.~R\'oza\'nski, {\em The residual gravity acceleration 
effect in the Poincar\'e dodecahedral space}, Astronomy and Astrophysics 502 (2009) 27 [11 pages]

\bibitem{SouHa} T.~Souradeep, A.~Hajian, {\em Statistical isotropy of CMB anisotropy from WMAP},
arXiv:astro-ph/0502248.

\bibitem{Sper} D.N.~Spergel, L.~Verde, H.V.~Peiris, E.~Komatsu, M.R.~Nolta, C.L.~Bennett, M.~Halpern, G.~Hinshaw, N.~Jarosik, A.~Kogut, M.~Limon, S.S.~Meyer, L.~Page, G.S.~Tucker, J.L.~Weiland, E.~Wollack, E.L.~Wright, {\em First year Wilkinson Microwave Anisotropy Probe (WMAP) observations: determination of cosmological parameters}, Astrophys. J. Suppl. 148 (2003) 175--194.

\bibitem{Teg} M.~Tegmark, A.~de Oliveira-Costa, A.~Hamilton,
{\em A high resolution foreground cleaned CMB map from WMAP}, Phys. Rev. D. 
68 (2003) 123523.

\bibitem{UKE} J.P.~Uzan, U.~Kirchner, Ulrich, G.F.R.~Ellis, 
{\em WMAP data and the curvature of space}, 
Mon. Not. Roy. Astron. Soc. 344 (2003) L65.

\bibitem{URLW} J.P.~Uzan, A.~Riazuelo, R.~Lehoucq, J.~Weeks, {\em 
Cosmic microwave background constraints on lens spaces.} 
Phys. Rev. D, 69 (2004), 043003, 4 pp.

\bibitem{WeGu} J.~Weeks, J.~Gundermann, {\em Dodecahedral topology
fails to explain quadrupole-octupole alignment}, Class. Quantum
Grav. 24 (2007) 1863--1866. 

\bibitem{WLU} J.~Weeks, R.~Lehoucq, J.P.~Uzan, 
{\em Detecting topology in a nearly flat spherical universe},
Class. Quant. Grav. 20 (2003) 1529--1542.


\end{thebibliography}
\end{document}